\documentclass{amsart}
\usepackage{amssymb,latexsym}

\usepackage{graphicx}
\usepackage{amscd}

\theoremstyle{plain}
\newtheorem{acknowledgement}{Acknowledgement}

\newtheorem{corollary}{Corollary}

\newtheorem{definition}{Definition}
\newtheorem{example}{Example}

\newtheorem{lemma}{Lemma}

\newtheorem{proposition}{Proposition}
\newtheorem{remark}{Remark}

\numberwithin{equation}{section}

\input{tcilatex}

\begin{document}
\title[Supersymmetric Yang-Mills theory]{Regularization of $2d$ supersymmetric Yang-Mills theory via non commutative
geometry.}
\author{K. Valavane$^{\ast }$.}
\address{Centre de Physique Th\'{e}orique \\
et Institut de Math\'{e}matique de Luminy\\
Campus de Luminy\\
171, avenue de Luminy\\
13009 Marseille.}
\email{valavane@cpt.univ-mrs.fr.}

\begin{abstract}
The non commutative geometry is a possible framework to regularize Quantum
Field Theory in a nonperturbative way. This idea is an extension of the
lattice approximation by non commutativity that allows to preserve
symmetries. The supersymmetric version is also studied and more precisely in
the case of the Schwinger model on supersphere [14]. This paper is a
generalization of this latter work to more general gauge groups.
\end{abstract}

\keywords{Fuzzy sphere, supersymmetric Yang-Mills, non commutative geometry.\\
$^{\ast }$Asistant Temporaire d'Enseignement et de Recherche \`{a}
l'Universit\'{e} de la M\'{e}diterran\'{e}e, Marseille.}
\maketitle

\section{Introduction}

Formally the quantization (in Feynman's point of view) of a field is
represented by a path integral , but this integral is not well defined [1].\
The lattice approximation was first proposed as a way to regularize this
integral but it does not preserve the Lorentz invariance. Snyder has
introduced non commutativity of the coordinates to conserve Lorentz symmetry
[23]. In this approach the space time is not a manifold but is decomposed in
cells of certain size (multiple of Planck constant). This approach
introduces a natural (UV) cut-off and it can be non perturbative. At least
in compact cases, this cut-off allows us to remove divergences. This fuzzy
approach [19, 20, 7, 15, 4, 14] of the regularization is exposed in the case
of sphere using Berezin quantization [2], the result is so-called fuzzy
sphere. In this framework, there are lot of works [12, 7, 21, 16, 4] which
are trying to include all the fields. But in the noncompact cases the (UV)
divergences can persist [5].

The fuzzy sphere is introduced by quantization of the symplectic structure
on the usual sphere. It replaces the commutative structure by non
commutative one and the quantum version of the symplectic reduction
introduces naturally the finiteness. The first step is the regularization of
a scalar field on the sphere [19, 10, 3]. The scalar field on fuzzy sphere
is just a matrix and the action (always invariant by $SO(3)$) is defined
using the trace on finite matrices.

Other field theories (spinors fields, gauge fields and topologically
nontrivial field configurations) are also defined on the fuzzy sphere [10,
7, 13, 4, 8, 14] and their regularization proved, thanks again to the
finiteness of the matrices. These constructions needed the non commutative
generalization of spinors, of the differential complex and of the
topologically nontrivial configurations. To know more about noncommutative
geometry and its applications, see [6]. In [11, 12], the definition of the
spinors (element of a bimodule) on fuzzy sphere, which allows to construct
Dirac operator and chiral operator. But the latter two didn't anticommute,
thus the previous assumption did not preserve the perfect analogy between
fuzzy sphere and the ordinary sphere. Another approach consists in using
supersymmetry [7]. The supersymmetric version of the fuzziness is very
similar to the ordinary case : fuzzy superspheres are finite supermatrices,
the scalars fields are just the even parts of the supermatrices (bosonic
submanifold) and spinors are the odd ones. In fact the scalar fields and
spinors are both contained in superscalars fields in a canonical way. One
can also construct gauge fields using a differential complex based on this
concept [15].

If we want to consider supersymmetric gauge theories, all these
constructions are constraint to be gauge invariant and supersymmetric
invariant. Using this idea, C. Klimcik constructs the supersymmetric
Schwinger model (analogue of the euclidean Maxwell field in two dimension)
on the ordinary sphere and on the fuzzy sphere [14]. He constructed an
suitable invariant supersymmetric differential complex based on the super
Lie algebra $sl(2,1)$ and its sub super Lie algebra $osp(2,1)$. He worked
out in detail the abelian case and we aim to study the non abelian case in
this paper.

For this prurpose, we conserve the general form of the action defining the
electrodynamic on the fuzzy supersphere but we need to modify the
differential complex to incorporate the non abelian case. At the commutative
limit, a long calculation allows us to describe (also in an original way)
the supersymmetric Yang-Mills theory on the ordinary sphere. This paper is
organized as follows :

In the section 2, we recall some preliminary notions that underlie our
framework : supersphere, symplectic reduction, quantization of the
supersphere, super Lie algebras $sl(2,1)$\thinspace and $osp(2,1)$ and
integration over the fermionic variables.

In the section 3, we construct the analogue of the supersymmetric
differential complex presented [14] in the bosonic case and we modify it to
include the non abelian case. Then we apply this construction of the
modified complex to the supersphere and fuzzy supersphere.

At the commutative limit, we obtain respectively the Schwinger model [14]
and the ordinary Yang-Mills theory on the supersphere.

Last section is devoted to conclusions.

\section{Preliminaries}

\subsection{Supersphere}

To perform easily the quantization of the sphere as a phase space, we use
the well known symplectic reduction of the complex plane $\mathbb{C}^{2}$ by
the group $U(1)$. We consider the complex plane $\mathbb{C}^{2}$ generated
by $\chi ^{\alpha }$, $\alpha =1,2$, with the following Poisson structure : 
\begin{equation}
\{f,g\}=\partial _{\chi ^{\alpha }}f\partial _{\overline{\chi }^{\alpha
}}g-\partial _{\overline{\chi }^{\alpha }}f\partial _{\chi ^{\alpha }}g.
\end{equation}
We call $\omega $, the 2-form underlying this symplectic structure. We
consider a moment map $J=\chi _{1}^{2}+\chi _{2}^{2}-1$ then we can
associate $U(1)$ vector field $X$ to $J$ by 
\begin{equation*}
dJ=\omega \left( X,.\right) .
\end{equation*}
In the submanifold $J^{-1}(0)$, the form $\omega $ is degenerated. We obtain
the standard 2-sphere $S^{2}$ with its symplectic structure by considering
the quotient on this submanifold $J^{-1}(0)$ by null-space of the 2-form $%
\omega $. In other words, the algebra of functions on the sphere consists of
functions on $\mathbb{C}^{2}$ with the property

\begin{equation}
\left\{ f,\chi _{1}^{2}+\chi _{2}^{2}-1\right\} =0.
\end{equation}
Moreover such two functions are equivalent if their difference is a function
of the following form $h\left( \chi _{1}^{2}+\chi _{2}^{2}-1\right) $. This
procedure is called the symplectic reduction with a moment map $\chi
_{1}^{2}+\chi _{2}^{2}-1$.

In analogy with the algebra of functions on the sphere defined by symplectic
reduction with respect to a moment map $\chi _{1}^{2}+\chi _{2}^{2}-1$ in
the complex plan $\mathbb{C}^{2}$, the algebra of functions $\mathcal{A}%
_{\infty }$\ on the supersphere is defined by (super) symplectic reduction
with respect to a moment map $\chi _{1}^{2}+\chi _{2}^{2}+\overline{a}a-1$
in the complex superplane $\mathbb{C}^{2,1}$, with additional fermionic or
grassmanian variables $\overline{a}$, $a$ [1]. The Poisson structure on \ $%
\mathbb{C}^{2,1}$ is the following 
\begin{equation}
\{f,g\}=\partial _{\chi ^{\alpha }}f\partial _{\overline{\chi }^{\alpha
}}g-\partial _{\overline{\chi }^{\alpha }}f\partial _{\chi ^{\alpha
}}g-\left( -1\right) ^{f}\left[ \partial _{a}f\partial _{\overline{a}%
}g+\partial _{\overline{a}}f\partial _{a}g\right]
\end{equation}
applied to coordinates, seen as functions, it gives 
\begin{equation}
\{\chi ^{\alpha },\overline{\chi }^{\beta }\}=\delta _{\alpha \beta },\qquad
\{a,\overline{a}\}=1,\qquad \alpha ,\beta =1,2.
\end{equation}
The following parametrization simplifies our work 
\begin{equation*}
z=\frac{\chi ^{1}}{\chi ^{2}},\qquad \overline{z}=\frac{\overline{\chi }^{1}%
}{\overline{\chi }^{2}},\qquad b=\frac{a}{\chi ^{2}},\qquad \overline{b}=%
\frac{\overline{a}}{\overline{\chi }^{2}}.
\end{equation*}
The Berezin integral on this algebra is written as, $e$ is the unit of $%
\mathcal{A}_{\infty }$%
\begin{equation}
I[f]=\frac{1}{2\pi i}\int \frac{d\overline{z}\,dz\,d\overline{b}\,db}{1+%
\overline{z}z+\overline{b}b}f,\qquad I[e]=1.
\end{equation}
This algebra is equipped with graded involution 
\begin{equation}
\left( \chi ^{\alpha }\right) ^{\ast }=\overline{\chi }^{\alpha },\qquad
\left( \overline{\chi }^{\alpha }\right) ^{\ast }=\chi ^{\alpha },\qquad
a^{\ast }=\overline{a},\qquad \overline{a}^{\ast }=-a.
\end{equation}
Like $sl\left( 2\right) $ on the sphere, the Lie superalgebra $sl(2,1)$ is
naturally represented on $\mathcal{A}_{\infty }$. First of all, we recall
that $sl(2,1)$ is generated by $R_{\pm },R_{3},\Gamma ,V_{\pm },D_{\pm }$
with the following super Lie algebra structure. We note $[.,.]_{+}$ the
anti-commutator. 
\begin{eqnarray*}
\lbrack R_{3},R_{\pm }] &=&\pm R_{\pm },\qquad \lbrack
R_{+},R_{-}]=2R_{3},\qquad \lbrack R_{i},\Gamma ]=0,\qquad i=+,-,3. \\
\lbrack D_{\pm },V_{\pm }] &=&0,\qquad \lbrack D_{\pm },V_{\mp }]_{+}=\pm 
\frac{1}{4}\Gamma ,\qquad \lbrack D_{\pm },D_{\pm }]_{+}=\mp \frac{1}{2}%
R_{\pm }, \\
\lbrack D_{\pm },D_{\mp }]_{+} &=&\frac{1}{2}R_{3},\qquad \lbrack V_{\pm
},V_{\pm }]_{+}=\pm \frac{1}{2}R_{\pm },\qquad \lbrack V_{\pm },V_{\mp
}]_{+}=-\frac{1}{2}R_{3}, \\
\lbrack R_{3},V_{\pm }] &=&\pm \frac{1}{2}V_{\pm },\qquad \lbrack R_{\pm
},V_{\pm }]=0,\qquad \lbrack R_{\pm },D_{\mp }]=D_{\pm }, \\
\left[ \Gamma ,V_{\pm }\right] &=&D_{\pm },\qquad \left[ \Gamma ,D_{\pm }%
\right] =V_{\pm }.
\end{eqnarray*}
The representation of $sl(2,1)$\ on $\mathcal{A}_{\infty }$ is defined in
the following way 
\begin{eqnarray}
V_{\pm }f &=&\left\{ v_{\pm },f\right\} ,\qquad \Gamma f=\left\{ \gamma
,f\right\} ,\qquad f\in \mathcal{A}_{\infty } \\
D_{\pm }f &=&\left\{ d_{\pm },f\right\} ,\qquad R_{3}f=\left\{
r_{i},f\right\} ,  \notag \\
R_{+}f &=&\left\{ r_{+},f\right\} ,\qquad R_{-}f=\left\{ r_{-},f\right\} . 
\notag
\end{eqnarray}
with respect to the following charges 
\begin{eqnarray}
r_{+} &=&\overline{\chi }^{1}\chi ^{2},\qquad r_{-}=\overline{\chi }^{2}\chi
^{1},\qquad r_{3}=\frac{1}{2}\left( \overline{\chi }^{1}\chi ^{1}-\overline{%
\chi }^{2}\chi ^{2}\right) ,\qquad \gamma =\overline{a}a+1 \\
2v_{+} &=&\overline{\chi }^{1}a+\overline{a}\chi ^{2},\qquad 2v_{-}=%
\overline{\chi }^{2}a-\overline{a}\chi ^{1},  \notag \\
2d_{+} &=&\overline{a}\chi ^{2}+\overline{\chi }^{1}a,\qquad 2d_{-}=-%
\overline{\chi }^{2}a-\overline{a}\chi ^{1}.  \notag
\end{eqnarray}
This representation is called Hamiltonian because it can be defined by the
super-Poisson structure (2.1). The derivatives $V_{\pm },D_{\pm },\Gamma
,R_{\pm },R_{3}$ can be also expressed in terms of the standard
supersymmetric derivatives $D$, $\overline{D}$, $Q$, $\overline{Q}$ in two
dimensions 
\begin{eqnarray*}
D &=&\partial _{b}+b\partial _{z},\qquad \overline{D}=\partial _{\overline{b}%
}+\overline{b}\partial _{\overline{z}}, \\
Q &=&\partial _{b}-b\partial _{z},\qquad \overline{Q}=\partial _{\overline{b}%
}-\overline{b}\partial _{\overline{z}}.
\end{eqnarray*}
We write the generators of $sl(2,1)$ using these four derivatives : 
\begin{eqnarray*}
D_{+} &=&\frac{1}{2}\left( D-\overline{z}\overline{D}\right) ,\qquad D_{-}=-%
\frac{1}{2}\left( \overline{D}+zD\right) , \\
V_{+} &=&\frac{1}{2}\left( Q+\overline{z}\overline{Q}\right) ,\qquad V_{-}=%
\frac{1}{2}\left( \overline{Q}-zQ\right) , \\
\Gamma &=&\overline{b}\partial _{\overline{b}}-b\partial _{b},\qquad R_{3}=%
\overline{z}\partial _{\overline{z}}-z\partial _{z}+\frac{1}{2}\overline{b}%
\partial _{\overline{b}}-\frac{1}{2}b\partial _{b}, \\
R_{+} &=&-\partial _{z}-\overline{z}^{2}\partial _{\overline{z}}-\overline{z}%
\overline{b}\partial _{\overline{b}},\qquad R_{-}=\partial _{\overline{z}%
}+z^{2}\partial _{z}-zb\partial _{b}.
\end{eqnarray*}

In the supersymmetric framework the Taylor expansion of the functions is
finite (because the nilpotency of the fermionic variables). An even element
writes 
\begin{equation}
f\left( \overline{z},z,\overline{b},b\right) =u(\overline{z},z)+b\psi (%
\overline{z},z)+\overline{b}\varphi (\overline{z},z)+\overline{b}bv(%
\overline{z},z)
\end{equation}
with $u$ and $v$ belong to the even part of $\mathbb{P}$, $\mathbb{P}$\ a
graded commutative algebra. And $\psi $ and $\phi $ in the odd one. Thus it
is globally even. It is same to the odd element. We recall the integration
on the fermionic variables 
\begin{eqnarray}
\int db &=&0,\quad \int d\overline{b}=0,\quad \int dbb=0,\quad \\
\int d\overline{b}\overline{b} &=&0,\qquad \int d\overline{b}\,db\,f(%
\overline{z},z,\overline{b},b)=\int d\overline{b}\left( \psi -\overline{b}%
v\right) =-v.  \notag
\end{eqnarray}

\subsection{Quantization of the supersphere}

In the previous part, we introduced the symplectic reduction because its
simplifies the quantization of the supersphere. Indeed, first we quantize
the superplane and we perform the quantum symplectic reduction [15]\footnote{%
There is an another way to quantize it, which is equivalent to the previous
one, using the representation theory of $sl(2,1)$ [7].}. As in quantum
mechanics, we transform the generators of the algebra in creation and
annihilation operators with the standard replacement 
\begin{equation}
\left\{ .,.\right\} \longrightarrow \frac{1}{h}[.,.]\text{ with }h\text{ is
real parameter.}
\end{equation}
Thus the generators $\overline{\chi }^{\alpha }$, $\chi ^{\alpha }$, $%
\overline{a}$, $a$ become operators verifying 
\begin{equation*}
\lbrack \chi ^{\alpha },\overline{\chi }^{\beta }]=h\delta _{\alpha \beta
},\qquad \lbrack a,\overline{a}]_{+}=h,\qquad \alpha ,\beta =1,2
\end{equation*}
and acting on a Hilbert space which is constructed as follows 
\begin{eqnarray*}
&&\overline{\chi }^{\alpha }\left| 0>\right. \text{ is an vector} \\
&&\overline{a}\left| 0>\right. \text{ is an another vector} \\
\chi ^{\alpha }\left| 0>\right. &=&0 \\
a\left| 0>\right. &=&0
\end{eqnarray*}
It means, one considers a vector (vacuum vector and the standard notation is 
$\left| 0>\right. $) and one constructs an irreducible representation of
this algebra. The space generated by this denumerable family of vectors is a
Hilbert space, called Fock space. The analogue of symplectic reduction with
moment map is just the restriction of the Hilbert space only to the vectors $%
\psi $ satisfying the constraint 
\begin{equation*}
\left( \overline{\chi }_{1}\chi _{1}+\overline{\chi }_{2}\chi _{2}+\overline{%
a}a-1\right) \psi =0
\end{equation*}
as in the previous section we define the quantized version of $\mathcal{A}%
_{\infty }$ by the operators $\widehat{f}$ which verify 
\begin{equation*}
\lbrack \widehat{f},\chi _{1}^{2}+\chi _{2}^{2}+\overline{a}a]=0.
\end{equation*}
Let us determine the dimension of the kernel of the operator $\chi
_{1}^{2}+\chi _{2}^{2}+\overline{a}a-1$. Let be $\psi $ an element of the
Fock space, it is easy to show that 
\begin{equation*}
\psi =\left( \overline{\chi }_{1}\right) ^{n_{1}}\left( \overline{\chi }%
_{2}\right) ^{n_{2}}\left| 0>\right. \text{ or }\psi =\left( \overline{\chi }%
_{1}\right) ^{n_{1}}\left( \overline{\chi }_{2}\right) ^{n_{2}}\overline{a}%
\left| 0>\right. \text{ with }n_{1},n_{2}\in \mathbb{N}
\end{equation*}
which implies that 
\begin{equation}
\left( \overline{\chi }_{1}\chi _{1}+\overline{\chi }_{2}\chi _{2}+\overline{%
a}a-1\right) \psi =Nh-1\text{ with }N\in \mathbb{N}.
\end{equation}
Thus the condition to fulfil (2.12) is that $h=\frac{1}{N}$ and in this
case, the dimension of the kernel of $\chi _{1}^{2}+\chi _{2}^{2}+\overline{a%
}a$ is just the number of possibilities to have $N=n_{1}+n_{2}$ or $%
n_{1}+n_{2}+1$, it is exactly $2N+1$. The each admissible value of the
parameter $h$ gives us a $(2N+1)$-dimensional subspace $H_{N}$ of the Fock
space \ and the deformed version of $\mathcal{A}_{\infty }$ is then $%
\mathcal{A}_{N}=M_{2N+1}(\mathbb{C})$. When $N\rightarrow \infty $ we have
the constant $h$ approaching $0$ and the algebra $\mathcal{A}_{N}$ tends to
the classical limit $\mathcal{A}_{\infty }$ [8]. The Hilbert space $H_{N}$
is graded $H_{N}=H_{eN}\oplus H_{oN}$ where $H_{eN}$ generated by bosonic
creation operators 
\begin{equation*}
\left( \overline{\chi }_{1}\right) ^{n_{1}}\left( \overline{\chi }%
_{2}\right) ^{n_{2}}\left| 0>\right. ,\qquad n_{1}+n_{2}=N
\end{equation*}
and $H_{oN}$ both bosonic and fermionic creation operators 
\begin{equation*}
\left( \overline{\chi }_{1}\right) ^{n_{1}}\left( \overline{\chi }%
_{2}\right) ^{n_{2}}\overline{a}\left| 0>\right. ,\qquad n_{1}+n_{2}+1=N.
\end{equation*}
The involution in $\mathcal{A}_{N}$ is defined exactly as in (2.6), $%
\mathcal{A}_{N}$ is also graded as follows [7] 
\begin{equation*}
\Phi =\left( 
\begin{array}{cc}
\phi _{R}\in M_{n+1}\left( \mathbb{C}\right) & \psi _{R}\in M_{n+1,n}\left( 
\mathbb{C}\right) \\ 
\psi _{L}\in M_{n,n+1}\left( \mathbb{C}\right) & \phi _{L}\in M_{n}\left( 
\mathbb{C}\right)
\end{array}
\right) \in \mathcal{A}_{N}
\end{equation*}
where even part is composed by diagonal blocks and odd by the off-diagonal
blocks. The integration over $\mathcal{A}_{N}$ is given by 
\begin{eqnarray*}
I[\Phi ] &=&STr[\Phi ],\qquad \Phi \in \mathcal{A}_{N}. \\
&=&Trace\left( \phi _{R}-\phi _{L}\right) .
\end{eqnarray*}
The relations of the super Lie bracket with the super-Poisson structure for $%
N\rightarrow \infty $ is given by 
\begin{equation}
\{X,Y\}=N[X,Y],\qquad X,Y\in \mathcal{A}_{N}.
\end{equation}
The graduation of the commutator depends on the graduation of the elements :
if $X$ and $Y$ are both odd, it is in fact the anti-commutator. The
representation defined by (2.7) on $\mathcal{A}_{\infty }$ is preserved by
quantization and becomes a representation on $\mathcal{A}_{N}$ in which we
replace the Poisson bracket by the graded commutator. In the ''quantum''
case, the action is defined by 
\begin{eqnarray}
V_{j}f &=&\left[ v_{\pm },f\right] ,\qquad \Gamma f=\left[ \gamma ,f\right] ,
\\
D_{\pm }f &=&\left[ d_{\pm },f\right] ,\qquad R_{3}f=\left[ r_{3},f\right] ,
\notag \\
R_{+}f &=&\left[ r_{+},f\right] ,\qquad R_{-}f=\left[ r_{-},f\right] . 
\notag
\end{eqnarray}
The explicit form of the supermatrices $r_{i},\gamma ,v_{\alpha },d_{\beta }$
are given in [14]. The representations of $sl(2,1)$ on $\mathcal{A}_{N}$ and 
$\mathcal{A}_{\infty }$ are completely reducible, their decompositions into
irreducible ones are the following 
\begin{equation*}
\mathcal{A}_{N}=\bigoplus_{j=0}^{N}j,\qquad \mathcal{A}_{\infty
}=\bigoplus_{j=0}^{\infty }j
\end{equation*}
where $j$ means the $sl(2,1)$ superspin of the representation, for more
details see [7, 14]. We recall that the quantization performed using the
representation theory of $sl(2,1)$ is just the approximation at level $N$ of 
$\mathcal{A}_{\infty }=\bigoplus_{j=0}^{\infty }j$ by $\mathcal{A}%
_{N}=\bigoplus_{j=0}^{N}j$ endowed with a new multiplication rule. It is
clear that at the limit $\mathcal{A}_{N}$ becomes $\mathcal{A}_{\infty }$,
for more details see [7].

In [14] C. Klimcik constructed an action of the supersymmetric gauge theory
for the finite $N$, at the limit it becomes the standard free supersymmetric
electrodynamic in the ordinary sphere. In the following section we construct
the modified differential complex that allows us to include the non abelian
case.

\section{Description of the modified differential complex}

\subsection{Bosonic case}

Firstly, we construct a differential complex on the ordinary sphere in an
invariant way and then we extend it to the supersphere [14]. The invariant
complex on the ordinary sphere is obtained by an another way in [16]. The
differential complex constructed in [14] can be seen as a supersymmetric
generalization of the following one.

\begin{definition}
A Poisson algebra $\mathcal{A}$ is an unital $\mathbb{C}$-algebra with a
Poisson structure compatible with the product $m:\mathcal{A\otimes A}%
\rightarrow \mathcal{A}$. 
\begin{equation*}
\{X,YZ\}=\{X,Y\}Z+Y\{X,Z\},\qquad X,Y,Z\in \mathcal{A}
\end{equation*}
$\mathcal{A}$ is equipped with a linear trace 
\begin{eqnarray*}
Trace &:&\mathcal{A}\rightarrow \mathbb{C} \\
Trace(e) &=&1\text{ where }e\text{ is\ the unit of }\mathcal{A}, \\
Trace(\{X,Y\}) &=&0,\qquad X,Y\in \mathcal{A}.
\end{eqnarray*}
\end{definition}

\begin{definition}
We say that $(\mathcal{A},\mathcal{G})$ is a double over a Poisson $\mathbb{C%
}$-algebra $\mathcal{A}$ if $\mathcal{G}$ is a Lie subalgebra of $\mathcal{A}
$ and a bilinear form $Trace\circ m$ restricted to $\mathcal{G}$ is
non-degenerated. In this case the bilinear form $Trace\circ m$ determines an
element $C^{\mathcal{G}}\in \mathcal{G}\otimes \mathcal{G}$ called a
quadratic Casimir element of the double $(\mathcal{A},\mathcal{G})$.
\end{definition}

These two definitions allow us to construct a invariant differential complex
on $\mathcal{A}$ by the following way : The complex $\Omega \left( \mathcal{A%
},\mathcal{G}\right) $ over the double $(\mathcal{A},\mathcal{G})$ is
defined as follows 
\begin{equation}
\Omega \left( \mathcal{A},\mathcal{G}\right) =\bigoplus_{i=0}^{3}\Omega
_{i}\left( \mathcal{A},\mathcal{G}\right)
\end{equation}
where 
\begin{eqnarray}
\Omega _{0}\left( \mathcal{A},\mathcal{G}\right) &=&\Omega _{3}\left( 
\mathcal{A},\mathcal{G}\right) =\left( \mathcal{A}\right) _{0}\equiv
e\otimes \mathcal{A} \\
\Omega _{1}\left( \mathcal{A},\mathcal{G}\right) &=&\Omega _{2}\left( 
\mathcal{A},\mathcal{G}\right) =\mathcal{G\otimes A}  \notag
\end{eqnarray}
We note $m$ the left regular action and $ad$ the adjoint action of $\mathcal{%
A}$ on itself. We explicit their actions 
\begin{eqnarray*}
ad\left( X\right) Y &=&\{X,Y\} \\
m\left( X\right) Y &=&XY
\end{eqnarray*}
Using Sweedler notation, we note formally $C^{\mathcal{G}}$ as $C_{1}^{%
\mathcal{G}}\otimes C_{2}^{\mathcal{G}}\in \mathcal{G}\otimes \mathcal{G}$.
Let us introduce now the coboundary operator 
\begin{equation}
\delta ^{\mathcal{G}}:\Omega _{i}\left( \mathcal{A},\mathcal{G}\right)
\longrightarrow \Omega _{i+1}\left( \mathcal{A},\mathcal{G}\right)
\end{equation}
which acts explicitly 
\begin{eqnarray*}
\delta ^{\mathcal{G}}\left( e\otimes X\right) &=&m\left( C_{1}^{\mathcal{G}%
}\right) e\otimes ad\left( C_{2}^{\mathcal{G}}\right) X,\qquad e\otimes X\in
\Omega _{0}\left( \mathcal{A},\mathcal{G}\right) \\
\delta ^{\mathcal{G}}\left( g\otimes X\right) &=&\left( ad\left( C_{1}^{%
\mathcal{G}}\right) \otimes ad\left( C_{2}^{\mathcal{G}}\right) +\frac{1}{2}%
d_{\mathcal{G}}\right) \left( g\otimes X\right) ,\qquad g\otimes X\in \Omega
_{1}\left( \mathcal{A},\mathcal{G}\right) \\
\delta ^{\mathcal{G}}\left( k\otimes Y\right) &=&e\otimes ad\left( k\right)
Y,\qquad k\otimes Y\in \Omega _{2}\left( \mathcal{A},\mathcal{G}\right) \\
\delta ^{\mathcal{G}}\left( e\otimes W\right) &=&0,\qquad e\otimes W\in
\Omega _{3}\left( \mathcal{A},\mathcal{G}\right) .
\end{eqnarray*}
with $d_{\mathcal{G}}$ the Dynkin index for the trace, which can be defined
by 
\begin{equation*}
Trace(XY)=\frac{1}{d_{\mathcal{G}}}Trace(ad\left( X\right) ad\left( Y\right)
)
\end{equation*}
We define also the associative graded product on this differential algebra
which is compatible with $\delta ^{\mathcal{G}}$%
\begin{equation}
\ast _{\mathcal{G}}:\Omega _{i}\left( \mathcal{A},\mathcal{G}\right) \otimes
\Omega _{j}\left( \mathcal{A},\mathcal{G}\right) \longrightarrow \Omega
_{i+j}\left( \mathcal{A},\mathcal{G}\right)
\end{equation}
The multiplication is given by the following table 
\begin{equation*}
\left( 
\begin{array}{ccccc}
\ast _{\mathcal{G}} & e\otimes X^{\prime } & g^{\prime }\otimes X^{\prime }
& k^{\prime }\otimes Z^{\prime } & e\otimes W^{\prime } \\ 
e\otimes X & m\otimes m & m\otimes m & m\otimes m & m\otimes m \\ 
g\otimes X & m\otimes m & ad\otimes m & \left( Trace\otimes Id\right) \left(
m\otimes m\right) & 0 \\ 
k\otimes Z & m\otimes m & \left( Trace\otimes Id\right) \otimes \left(
m\otimes m\right) & 0 & 0 \\ 
e\otimes W & m\otimes m & 0 & 0 & 0
\end{array}
\right)
\end{equation*}
Finally we define a map, called Hodge triangle, which is the analogue of the
Hodge star. It is just the identification between 2-forms and 1-forms,
between 0-forms and 3-forms and we denote it $\lhd $. This presentation is
just the application of the one constructed in the supersymmetric case in
[14] to the bosonic case. In [16] C. Klimcik showed that this complex
applied to $\mathcal{A}=C^{\infty }(S^{2})$ and $\mathcal{G}=su(2)$ is
isomorphic to another one constructed with the de Rham complex of the
2-sphere [16]. Now we recall it : 
\begin{equation}
\omega =\underset{i=0}{\overset{3}{\oplus }}\omega _{i}
\end{equation}
with 
\begin{eqnarray}
\omega _{0} &=&\Omega _{0}\oplus \left\{ 0\right\} ,\qquad \omega
_{1}=\Omega _{1}\oplus \Omega _{2} \\
\omega _{2} &=&\Omega _{2}\oplus \Omega _{1},\qquad \omega _{3}=\left\{
0\right\} \oplus \Omega _{0}.  \notag
\end{eqnarray}
We note $\Omega _{i}$, the space of $i$-forms in the usual de Rham complex, $%
d$ the de Rham differential operator and $\ast $ the usual Hodge operator.
The coboundary operator on $\omega $ is defined as follows 
\begin{eqnarray}
\delta &\equiv &d\oplus \ast d\ast \text{ on }\omega _{0},\omega _{2},\omega
_{3} \\
\delta \left( V\oplus v\right) &=&\left( dV+v\right) \oplus \ast d\ast
v,\qquad V\oplus v\in \Omega _{1}\oplus \Omega _{2}  \notag
\end{eqnarray}
We recall the definition of the Hodge triangle for this complex [16] 
\begin{eqnarray}
\triangleleft \phi &=&\phi ,\qquad \phi \in \omega _{0} \\
\triangleleft \left( V\oplus v\right) &=&v\oplus V,\qquad V\oplus v\in
\omega _{1} \\
\triangleleft \left( v\oplus V\right) &=&V\oplus v,\qquad v\oplus V\in
\omega _{2}  \notag \\
\triangleleft \Phi &=&\Phi ,\qquad \Phi \in \omega _{3}.
\end{eqnarray}
We define the integral of a 3-form by 
\begin{equation}
Int(\Phi )=\int_{S^{2}}\ast \Phi
\end{equation}
where $\Phi $ is seen as a 0-form of the de Rham complex. Now we can write
an action 
\begin{equation}
S(A)=Int(\delta A.\triangleleft \delta A)+Int(A.\delta A)
\end{equation}
where $A=V\oplus v\in \omega _{1}$ and $A^{\dagger }=A$. This action is
gauge invariant under the gauge transformation 
\begin{equation}
A\longrightarrow A+\delta \phi ,\qquad \phi \in \omega _{0}.
\end{equation}
Explicitly this action is written 
\begin{equation}
S(V,v)=\int_{S^{2}}dV\wedge \ast dV+\int_{S^{2}}d\ast v\wedge \ast d\ast v
\end{equation}
and the gauge transformation is written as follows 
\begin{equation*}
V\longrightarrow V+d\phi ,\qquad v\longrightarrow v.
\end{equation*}
The second term of the action (3.12) does not violate the gauge invariance
and it is useful to separate the fields $V$ and $v$ in the action. The first
term is the pure electrodynamic plus free massless scalar on $S^{2}$. Its
interaction with a scalar matter field $\Phi \in \omega _{0}^{\mathbb{C}}$
is described by [16] 
\begin{equation}
S_{m}(A,\Phi )=Int\left( \left( \delta \Phi -iA\Phi \right) .\triangleleft 
\overline{\left( \delta \Phi -iA\Phi \right) }\right)
\end{equation}
the bare means ordinary complex conjugation. In terms of fields $V,v$, the
matter action becomes 
\begin{equation}
S_{m}(\Phi ,V,v)=\int \left( d\Phi -iV\Phi \right) \wedge \ast \overline{%
\left( d\Phi -iV\Phi \right) }+\int \left( v\wedge \ast v\right) \Phi 
\overline{\Phi }
\end{equation}
This action is the standard interaction of the complex scalar matter with $%
U(1)$ gauge field but the second term is a new one. With the convenient
constraint which respects the gauge invariance, we can suppress $v$. Thus we
are able to construct the non commutative version without extra propagating
fields.

This complex can be also viewed as a subcomplex of the de Rham complex on $%
SU(2)$. Forms in this subcomplex are characterized by their invariance with
respect to $U(1)$ subgroup of $SU(2)$. So they can be interpreted as objects
living on $S^{2}(=SU(2)/U(1))$ [16]. The $SU(2)$ covariant formalism for the
complex $\omega $ is exactly our complex $\Omega \left( \mathcal{A},\mathcal{%
G}\right) $ in case of $\mathcal{A}=C^{\infty }(S^{2})$ and $\mathcal{G}%
=su(2)$.

We note $R_{i}$, with $i=+,-,3$, the generators of the Lie algebra $su(2)$
with the following relations 
\begin{equation}
\left[ R_{+},R_{-}\right] =2R_{3},\qquad \left[ R_{3},R_{\pm }\right] =\pm
R_{\pm }
\end{equation}
It is easy to show that in this case 
\begin{eqnarray}
C^{\mathcal{G}} &=&R_{+}\otimes R_{-}+R_{-}\otimes R_{+}+\frac{1}{2}%
R_{3}\otimes R_{3} \\
d_{\mathcal{G}} &=&-2  \notag \\
\Omega _{0}\left( \mathcal{A},\mathcal{G}\right) &=&\Omega _{3}\left( 
\mathcal{A},\mathcal{G}\right) =e\otimes \mathcal{A}=\Omega _{0}  \notag \\
\Omega _{1}\left( \mathcal{A},\mathcal{G}\right) &=&\Omega _{2}\left( 
\mathcal{A},\mathcal{G}\right) =\mathcal{G}\otimes \mathcal{A=}\Omega
_{0}\oplus \Omega _{0}\oplus \Omega _{0}  \notag
\end{eqnarray}
with $r_{i}$ the following charges corresponding to vector field $R_{i}$%
\begin{equation}
r_{+}=\overline{\chi }^{1}\chi ^{2}=\frac{\overline{z}}{1+\overline{z}z}%
,\qquad r_{-}=\overline{\chi }^{2}\chi ^{1}=\frac{z}{1+\overline{z}z},\qquad
r_{3}=\left( \overline{\chi }^{1}\chi ^{1}-\overline{\chi }^{2}\chi
^{2}\right) =\frac{1}{2}\frac{1-\overline{z}z}{1+\overline{z}z}.
\end{equation}
We recall that $C^{\infty }(S^{2})$ is the $0$-forms in the de Rham complex.
Let us to note 
\begin{eqnarray}
\varphi &\equiv &e\otimes \varphi \in \Omega _{0}\cong C^{\infty }(S^{2}), \\
A_{i} &\equiv &R_{i}\otimes A_{i}\in \Omega _{1}\left( \mathcal{A},\mathcal{G%
}\right) \cong C^{\infty }(S^{2})\oplus C^{\infty }(S^{2})\oplus C^{\infty
}(S^{2}),  \notag \\
a_{i} &\equiv &R_{i}\otimes a_{i}\in \Omega _{3}\left( \mathcal{A},\mathcal{G%
}\right) \cong C^{\infty }(S^{2})\oplus C^{\infty }(S^{2})\oplus C^{\infty
}(S^{2}),  \notag \\
\Phi &\equiv &e\otimes \Phi \in \Omega _{0}.\cong C^{\infty }(S^{2})  \notag
\end{eqnarray}
So the multiplication becomes explicitly 
\begin{equation}
m\otimes m\left( e\otimes \varphi \right) \left( R_{i}\otimes A_{i}\right)
=R_{i}\otimes \varphi A_{i}\equiv \varphi A_{i},
\end{equation}
in this way we obtain the complete table 
\begin{eqnarray*}
\varphi \ast _{\mathcal{G}}\psi &=&\varphi \psi ,\qquad \varphi \ast _{%
\mathcal{G}}A_{i}=\varphi A_{i},\qquad \varphi ,\psi \in \Omega _{0},A\in
\Omega _{1} \\
\varphi \ast _{\mathcal{G}}b_{i} &=&\varphi b_{i},\qquad \varphi \ast _{%
\mathcal{G}}\Phi =\varphi \Phi ,\qquad \varphi \in \Omega _{0},b\in \Omega
_{2} \\
A\ast _{\mathcal{G}}B &=&\left(
A_{3}B_{+}-A_{+}B_{3},A_{3}B_{-}-A_{-}B_{3},2\left(
A_{+}B_{-}-A_{-}B_{+}\right) \right) ,\qquad A,B\in \Omega _{1} \\
A\ast _{\mathcal{G}}a &=&A_{+}a_{+}+A_{-}a_{-}+A_{3}a_{3},\qquad A\in \Omega
_{1},a\in \Omega _{2}.
\end{eqnarray*}
Now I explicit the coboundary operator 
\begin{eqnarray}
\delta ^{\mathcal{G}}\left( e\otimes \varphi \right) &=&\left( R_{-},\varphi
,R_{+}\varphi ,R_{3}\varphi \right)  \notag \\
\delta ^{\mathcal{G}}\left( r_{i}\otimes A_{i}\right)
&=&(-R_{-}A_{3}-R_{3}A_{+}-A_{+},  \notag \\
&&R_{+}A_{3}+R_{3}A_{-}-A_{-},  \notag \\
&&2R_{-}A_{-}-2R_{+}A_{+}-A_{3},)  \notag \\
\delta ^{\mathcal{G}}\left( r_{i}\otimes a_{i}\right)
&=&R_{+}A_{+}+R_{-}A_{-}+R_{3}A_{3}  \notag
\end{eqnarray}
The identification between the two descriptions in the case of the 1-forms
is 
\begin{eqnarray}
A_{+} &=&-iV-i\overline{z}^{2}\overline{V}+\frac{\overline{z}}{1+\overline{z}%
z}\ast v \\
A_{-} &=&i\overline{V}+iz^{2}V+\frac{z}{1+\overline{z}z}\ast v  \notag \\
A_{3} &=&i\overline{z}\overline{V}-izV+\frac{1}{2}\frac{1-\overline{z}z}{1+%
\overline{z}z}\ast v  \notag
\end{eqnarray}
where $Vdz+\overline{V}d\overline{z}$ and $v$ are de Rham 1-form and 2-form
respectively with $\overline{V}$, $V$ functions on $S^{2}$ verify 
\begin{equation*}
\int_{S^{2}}d\overline{z}dz\overline{V}V<\infty .
\end{equation*}
The first integral of the matter action (3.16) is the standard interaction
of the complex scalar matter with $U(1)$ gauge field. We impose certain
constraint to eliminate the second term [16], using the isomorphism (3.22)
the constraint is 
\begin{equation}
r_{+}A_{-}+r_{-}A_{+}+r_{3}A_{3}=0.
\end{equation}
Using (3.19) and (3.22) it is easy to show that this constraint eliminate $v$%
. In the invariant description this constraint is written 
\begin{equation}
C^{\mathcal{G}}\ast _{\mathcal{G}}\lhd A=0,\text{\qquad }A\in \Omega
_{1}\left( \mathcal{A},\mathcal{G}\right)
\end{equation}
It is important to note that constraint (3.24) is gauge and $su(2)$
invariant. In this constraint $C^{\mathcal{G}}$ is viewed as a 2-form. All
these constructions are extensible to the fuzzy sphere [20]. Briefly we
recall it 
\begin{eqnarray}
\mathcal{A}_{N} &=&M_{N}\left( \mathbb{C}\right) , \\
\left\{ f,g\right\} &=&N\left[ f,g\right] ,\qquad f,g\in \mathcal{A}_{N} 
\notag
\end{eqnarray}
\begin{eqnarray}
\widetilde{\omega }_{0} &=&\mathcal{A}_{N},\qquad \widetilde{\omega }_{1}=%
\mathcal{A}_{N}\otimes \mathbb{C}^{3} \\
\widetilde{\omega }_{2} &=&\mathcal{A}_{N}\otimes \mathbb{C}^{3},\qquad 
\widetilde{\omega }_{3}=\mathcal{A}_{N}.
\end{eqnarray}
The product between forms is defined as in (3.4). The coboundary operator is 
\begin{eqnarray}
\delta \left( \varphi \right) &=&\left( \left[ r_{-},\varphi \right] ,\left[
r_{+},\varphi \right] ,\left[ r_{3},\varphi \right] \right) \\
\delta \left( A\right) &=&(-\left[ r_{-},A_{3}\right] -\left[ r_{3},A_{+}%
\right] -A_{+},  \notag \\
&&\left[ r_{+},A_{3}\right] +\left[ r_{3},A_{-}\right] -A_{-},  \notag \\
&&2r_{-},A_{-}-2r_{+},A_{+}-A_{3},)  \notag \\
\delta \left( a\right) &=&\left[ r_{+},A_{+}\right] +\left[ r_{-},A_{-}%
\right] +\left[ r_{3},A_{3}\right]  \notag
\end{eqnarray}
with quantized charges of the Hamiltonian vectors $r_{i}$ which are
operators defined as in (3.19). The $Int$ (3.11) is becomes $\frac{1}{N+1}%
Trace$ in the noncommutative case. Then the action is written 
\begin{equation}
S_{N}(A)=\frac{1}{N+1}Trace\left( F.\triangleleft F+A.\delta A\right)
\end{equation}
The natural way to consider the non abelian case is to introduce a new gauge
transformation law of the 1-fields as follows 
\begin{equation}
A\longrightarrow UAU^{-1}+\delta UU^{-1},\qquad U\in \mathcal{U}_{N}\left( 
\mathbb{C}\right)
\end{equation}
To preserve to gauge invariance of the action (3.29), we modify it as
follows 
\begin{equation}
S_{N}(A)=\frac{1}{N+1}Trace\left( F.\triangleleft F++A.\delta A+\frac{2}{3}%
A.A.A\right) ,\qquad F=\delta A+A.A.
\end{equation}
The analogue of the constraint (3.23) for this action is 
\begin{equation}
r.\triangleleft A+\triangleleft A.r+A.\triangleleft A=0
\end{equation}
In a invariant description 
\begin{equation}
C^{\mathcal{G}}\ast _{\mathcal{G}}\lhd A+\lhd A\ast _{\mathcal{G}}C^{%
\mathcal{G}}+\frac{1}{N}\left( \triangleleft A\ast _{\mathcal{G}}A\right) =0
\end{equation}
In the commutative limit, $N\rightarrow \infty $, terms $A.A$ and $A.A.A$
vanished and we obtain (3.29). Thus one obtains the noncommutative version
of the scalar Maxwell theory on the sphere [16]. Now we can naturally
incorporate the Yang-Mills system into this framework. It implies that $%
\mathcal{A}_{N}$ should be replaced by $\mathcal{A}_{N}^{\prime }\equiv 
\mathcal{A}_{N}\otimes M_{n}(\mathbb{C})$. The gauge group $\mathcal{U}$ can
be viewed as the unitary elements of $\mathcal{A}_{N}\otimes M_{n}(\mathbb{C}%
)$. Since $\mathcal{A}_{N}=M_{N}(\mathbb{C)}$, we have $\mathcal{U}=\mathcal{%
U}_{nN}(\mathbb{C})$. In this formalism the only thing to modify is the
coboundary operator, we recall that coboundary operator is defined using the
charges $r_{i}$, therefore the modification concern them. We set 
\begin{eqnarray*}
\delta ^{\prime }\left( \phi \otimes m\right) &=&\left( \left[ r_{-},\phi %
\right] \otimes m,\left[ ,r_{+},\phi \right] \otimes m,\left[ r_{3},\phi %
\right] \otimes m\right) .\qquad \phi \otimes m\in \mathcal{A}_{N}^{\prime }
\\
\delta ^{\prime }\left( A_{i}\otimes n_{i}\right) &=&(-\left[ r_{-},A_{3}%
\right] \otimes n_{3}-\left[ r_{3},A_{+}\right] \otimes n_{+}-A_{+}\otimes
n_{+}, \\
&&\left[ r_{+},A_{3}\right] \otimes n_{3}+\left[ r_{3},A_{-}\right] \otimes
n_{-}-A_{-}\otimes n_{-}, \\
&&2\left[ r_{-},A_{-}\right] \otimes n_{-}-2\left[ r_{+},A_{+}\right]
\otimes n_{3}-A_{3}\otimes n_{3},). \\
\delta ^{\mathcal{G}}\left( a_{i}\otimes p_{i}\right) &=&\left[ r_{+},a_{+}%
\right] \otimes p_{+}+\left[ r_{-},a_{-}\right] \otimes p_{-}+\left[
r_{3},A_{3}\right] \otimes p_{3}. \\
\delta ^{\prime }\left( \Phi \otimes q\right) &=&0.\qquad \Phi \otimes q\in 
\mathcal{A}_{N}^{\prime }
\end{eqnarray*}
with $\left( A_{i}\otimes n_{i}\right) ,\left( a_{i}\otimes p_{i}\right) \in 
\mathcal{A}_{N}^{\prime }\otimes \mathbb{C}^{3}$. In the same manner, the
Casimir element $C^{\mathcal{G}}$ becomes 
\begin{equation}
C^{\mathcal{G}}=C_{1}^{\mathcal{G}}\otimes C_{2}^{\mathcal{G}}\otimes 
\mathbb{I}_{n}
\end{equation}
The gauge invariant analogue of the constraint (3.24) is 
\begin{equation}
C^{\mathcal{G}}\ast _{\mathcal{G}}\lhd A+\lhd A\ast _{\mathcal{G}}C^{%
\mathcal{G}}+\frac{1}{N}\triangleleft A\ast _{\mathcal{G}}A=0
\end{equation}
This previous bosonic work allowed us to understand the way to incorporate
abelian and non abelian theories in a same framework. Now we introduce the
modified differential complex which will be used in the supersymmetric
framework.

\subsection{Description of the modified differential complex}

Now we will construct a differential complex on the supersphere and the
supergauge abelian and non abelian theories on it. This complex is slightly
different from the complex constructed in [14] in order to incorporate the
non abelian theory on the supersphere. For a general propose, we consider $%
\mathcal{A}$, a $\mathbb{Z}_{2}$-graded unital $\mathbb{C}$-algebra with a
super-Poisson structure and $\mathcal{A}^{\prime }=\mathcal{A}\otimes
M_{n}\left( \mathbb{C}\right) $. The product on $\mathcal{A}^{\prime }$ is 
\begin{equation*}
\left( X\otimes m\right) \cdot \left( Y\otimes n\right) =XY\otimes mn,\qquad
X\otimes m,Y\otimes n\in \mathcal{A}^{\prime }
\end{equation*}
Now we define a bilinear map on $\mathcal{A}^{\prime }$ as follows 
\begin{eqnarray*}
\mathcal{A}^{\prime }\times \mathcal{A}^{\prime } &\longrightarrow &\mathcal{%
A}^{\prime } \\
\{X\otimes m,Y\otimes n\} &=&\{X,Y\}\otimes mn,\qquad X\otimes m,Y\otimes
n\in \mathcal{A}^{\prime }
\end{eqnarray*}
with $\{.,.\}$ the super Poisson structure on $\mathcal{A}$ compatible with
the product on $\mathcal{A}$. The restriction of this map on the subalgebra $%
\mathcal{A\equiv A}^{\prime }\otimes \mathbb{I}_{n}$ is a super Poisson
structure compatible with the product. But the map is not a super Poisson on 
$\mathcal{A}\otimes M_{n}\left( \mathbb{C}\right) $. Before giving our
definitions, let us list those appropriated for the abelian case.

\begin{definition}[14]
$\mathcal{A}$ is a $\mathbb{Z}_{2}$-graded unital $\mathbb{C}$-algebra with
a super-Poisson structure $\{.,.\}$ compatible with the product and equipped
with a linear supertrace 
\begin{eqnarray*}
STrace &:&\mathcal{A}\rightarrow \mathbb{C} \\
STrace(e) &=&1\text{ where }e\text{ is\ the unit of }\mathcal{A}. \\
STrace(\{X,Y\}) &=&0,\qquad X,Y\in \mathcal{A}
\end{eqnarray*}
\end{definition}

\begin{definition}[14]
We say that $(\mathcal{A},\mathcal{G})$ is a supersymmetric double over a
super-Poisson $\mathbb{P}$-algebra $\mathcal{A}$ with $\mathbb{P}$\ a graded
commutative algebra, if $\mathcal{G}=\mathcal{G}_{0}\oplus \mathcal{G}_{1}$
is a super-Lie subalgebra of $\mathcal{A}$ and a bilinear form $STrace\circ
m $ restricted to $\mathcal{G}$ is non-degenerated. In this case the
bilinear form $STrace\circ m$ determines an element $C^{\mathcal{G}}\in 
\mathcal{G}\otimes \mathcal{G}$ called a quadratic Casimir element of the
double $(\mathcal{A},\mathcal{G})$.
\end{definition}

\begin{example}
The algebras $\mathcal{A}_{\infty }$ and $\mathcal{A}_{N}$ with these
super-Poisson structures (2.3) (2.13) and Berezin integral or supertrace on
the matrices. For $\mathcal{G}$\ we take naturally $\mathcal{G}$ imbedded as
super-Lie subalgebra on $\mathcal{A}$ via (2.8 ).
\end{example}

\begin{definition}[14]
We say that $\left( \mathcal{A},\mathcal{G},\mathcal{H}\right) $ is a
supersymmetric triple, if it exists a subspace $\mathcal{H}$ of $\mathcal{A}$
such that

1) $\mathcal{H}$ is a super-Lie subalgebra of $\mathcal{G}$,

2) $\left( \mathcal{A},\mathcal{H}\right) $ is the supersymmetric double
with the Casimir element $C^{\mathcal{H}}\in \mathcal{H\otimes H}$,
coboundary $\delta ^{\mathcal{H}}$ and product $\ast _{\mathcal{H}}$,

3) An element $C\equiv C^{\mathcal{G}}-C^{\mathcal{H}}$ fulfils $m(C)\in 
\mathbb{C}e$,

4) $ad\left( \mathcal{H}^{\perp }\mathcal{\otimes H}^{\perp }\right) \subset 
\mathcal{H}$ where $\mathcal{H}^{\perp }$ is an orthogonal complement of $%
\mathcal{H}$\ in $\mathcal{G}$ with respect to $STrace\circ m$.
\end{definition}

We write $\left( \mathcal{A}\right) _{0}$ $\left( \left( \mathcal{A}\right)
_{1}\right) $\ is even (odd) part with respect to the sum of grading of $%
\mathcal{A}$\ and of $\mathbb{P}$. $\mathbb{P}$ can be Grassmanian algebras
or \ graded matrix algebras. In the non abelian case any element of $%
\mathcal{A}\otimes M_{n}(\mathbb{C})$ is a matrix in which each component is
a element $\mathcal{A}$ with respect to previous graduation. We note $m$ the
left regular action and $ad$ the adjoint action of $\mathcal{A}$ on itself.
We have 
\begin{eqnarray}
ad\left( X\right) Y &=&(-1)^{X}\{X,Y\} \\
m\left( X\right) Y &=&(-1)^{Y}XY
\end{eqnarray}
\begin{eqnarray}
\widetilde{ad}\left( X\otimes n\right) \left( Y\otimes p\right) &\equiv
&(-1)^{X}\{X,Y\}\otimes np \\
\widetilde{m}\left( X\otimes n\right) \left( Y\otimes p\right) &\equiv
&(-1)^{Y}XY\otimes np
\end{eqnarray}
We call modified Casimir the following element which is written formally as 
\begin{equation}
\widetilde{C}^{\mathcal{G}}\equiv C_{1}^{\mathcal{G}}\otimes C_{2}^{\mathcal{%
G}}\otimes \mathbb{I}_{n}\in \mathcal{G}\otimes \mathcal{G}\otimes \mathbb{I}%
_{n}
\end{equation}
We note $d_{\mathcal{G}}$, analogue of the Dynkin index for the supertrace,
which can be defined by 
\begin{equation}
STrace(XY)=\frac{1}{d_{\mathcal{G}}}STrace(ad\left( X\right) ad\left(
Y\right) )
\end{equation}

\begin{proposition}
The modified complex $\widetilde{\Omega }$ over $\mathcal{A}^{\prime }=%
\mathcal{A}\otimes M_{n}(C)$ is defined as follows 
\begin{equation*}
\widetilde{\Omega }\left( \mathcal{A}^{\prime },\mathcal{G}\right)
=\bigoplus_{i=0}^{3}\widetilde{\Omega }_{i}\left( \mathcal{A}^{\prime },%
\mathcal{G}\right)
\end{equation*}
where 
\begin{eqnarray*}
\widetilde{\Omega }_{0}\left( \mathcal{A}^{\prime },\mathcal{G}\right) &=&%
\widetilde{\Omega }_{3}\left( \mathcal{A}^{\prime },\mathcal{G}\right)
=e\otimes \left( \mathcal{A}\right) _{0}\otimes M_{n}(\mathbb{C}) \\
\widetilde{\Omega }_{1}\left( \mathcal{A}^{\prime },\mathcal{G}\right) &=&%
\widetilde{\Omega }_{2}\left( \mathcal{A}^{\prime },\mathcal{G}\right)
=\left( \mathcal{G}_{0}\otimes \left( \mathcal{A}\right) _{0}\otimes M_{n}(%
\mathbb{C})\right) \oplus \left( \mathcal{G}_{1}\otimes \left( \mathcal{A}%
\right) _{1}\otimes M_{n}(\mathbb{C})\right)
\end{eqnarray*}
Let us introduce now the coboundary operator 
\begin{eqnarray*}
\widetilde{\delta }^{\mathcal{G}} &:&\widetilde{\Omega }_{i}\left( \mathcal{A%
}^{\prime },\mathcal{G}\right) \longrightarrow \widetilde{\Omega }%
_{i+1}\left( \mathcal{A}^{\prime },\mathcal{G}\right) \\
\widetilde{\delta }^{\mathcal{G}}\left( e\otimes X\otimes n\right)
&=&(-1)^{C_{2}^{\mathcal{G}}}m\left( C_{1}^{\mathcal{G}}\right) e\otimes
ad\left( C_{2}^{\mathcal{G}}\right) X\otimes n, \\
\widetilde{\delta }^{\mathcal{G}}\left( g\otimes Y\otimes n\right)
&=&ad\left( C_{1}^{\mathcal{G}}\right) \left( g\right) \otimes ad\left(
C_{2}^{\mathcal{G}}\right) \left( Y\right) \otimes n+\frac{1}{2}d_{\mathcal{G%
}}\left( g\otimes Y\otimes n\right) , \\
\widetilde{\delta }^{\mathcal{G}}\left( k\otimes Z\otimes n\right)
&=&(-1)^{k}e\otimes ad\left( k\right) Z\otimes n, \\
\widetilde{\delta }^{\mathcal{G}}\left( e\otimes W\otimes n\right) &=&0,
\end{eqnarray*}
and the associative product between the forms compatible with $\widetilde{%
\delta }^{\mathcal{G}}$ 
\begin{equation}
\widetilde{\ast }_{\mathcal{G}}:\widetilde{\Omega }_{i}\left( \mathcal{A}%
^{\prime },\mathcal{G}\right) \otimes \widetilde{\Omega }_{j}\left( \mathcal{%
A}^{\prime },\mathcal{G}\right) \longrightarrow \widetilde{\Omega }%
_{i+j}\left( \mathcal{A}^{\prime },\mathcal{G}\right)
\end{equation}
is given by the following table 
\begin{equation*}
\left( 
\begin{array}{ccccc}
\widetilde{\ast }_{\mathcal{G}} & e\otimes X^{\prime } & g^{\prime }\otimes
X^{\prime } & k^{\prime }\otimes Z^{\prime } & e\otimes W^{\prime } \\ 
e\otimes X & \widetilde{m}\otimes \widetilde{m} & \widetilde{m}\otimes 
\widetilde{m} & \widetilde{m}\otimes \widetilde{m} & \widetilde{m}\otimes 
\widetilde{m} \\ 
g\otimes X & \widetilde{m}\otimes \widetilde{m} & \widetilde{ad}\otimes 
\widetilde{m} & \left( STrace\otimes Id\right) \otimes \left( \widetilde{m}%
\otimes \widetilde{m}\right) & 0 \\ 
k\otimes Z & \widetilde{m}\otimes \widetilde{m} & \left( STrace\otimes
Id\right) \otimes \left( \widetilde{m}\otimes \widetilde{m}\right) & 0 & 0
\\ 
e\otimes W & \widetilde{m}\otimes \widetilde{m} & 0 & 0 & 0
\end{array}
\right)
\end{equation*}
\end{proposition}

\begin{remark}
The complex of Klimcik complex [14] is exactly the previous one in the case $%
\mathcal{A}^{\prime }=\mathcal{A}$. We obtain it from the modified
differential complex in a natural way 
\begin{equation*}
\widetilde{C}^{\mathcal{G}}\equiv C_{1}^{\mathcal{G}}\otimes C_{2}^{\mathcal{%
G}}\in \mathcal{G}\otimes \mathcal{G}
\end{equation*}
and 
\begin{equation*}
\Omega \left( \mathcal{A},\mathcal{G}\right) =\bigoplus_{i=0}^{3}\Omega
_{i}\left( \mathcal{A},\mathcal{G}\right)
\end{equation*}
where 
\begin{eqnarray*}
\Omega _{0}\left( \mathcal{A},\mathcal{G}\right) &=&\Omega _{3}\left( 
\mathcal{A},\mathcal{G}\right) =e\otimes \left( \mathcal{A}\right) _{0} \\
\Omega _{1}\left( \mathcal{A},\mathcal{G}\right) &=&\Omega _{2}\left( 
\mathcal{A},\mathcal{G}\right) =\mathcal{G}_{0}\otimes \left( \mathcal{A}%
\right) _{0}\oplus \mathcal{G}_{1}\otimes \left( \mathcal{A}\right) _{1}.
\end{eqnarray*}
The coboundary operator is 
\begin{eqnarray*}
\delta ^{\mathcal{G}} &:&\Omega _{i}\left( \mathcal{A},\mathcal{G}\right)
\longrightarrow \Omega _{i+1}\left( \mathcal{A},\mathcal{G}\right) \\
\delta ^{\mathcal{G}}\left( e\otimes X\right) &=&\left( -1\right) ^{C_{2}^{%
\mathcal{G}}}m\left( C_{1}^{\mathcal{G}}\right) e\otimes ad\left( C_{2}^{%
\mathcal{G}}\right) X,\qquad e\otimes X\in \Omega _{0} \\
\delta ^{\mathcal{G}}\left( g\otimes X\right) &=&\left( ad\left( C_{1}^{%
\mathcal{G}}\right) \otimes ad\left( C_{2}^{\mathcal{G}}\right) +\frac{1}{2}%
d_{\mathcal{G}}\right) \left( g\otimes X\right) ,\qquad g\otimes X\in \Omega
_{1} \\
\delta ^{\mathcal{G}}\left( k\otimes Y\right) &=&\left( -1\right)
^{k}e\otimes ad\left( k\right) Y,\qquad k\otimes Y\in \Omega _{2} \\
\delta ^{\mathcal{G}}\left( e\otimes W\right) &=&0.\qquad e\otimes W\in
\Omega _{3}
\end{eqnarray*}
and the associative product between the forms 
\begin{equation}
\ast _{\mathcal{G}}:\Omega _{i}\left( \mathcal{A},\mathcal{G}\right) \otimes
\Omega _{j}\left( \mathcal{A},\mathcal{G}\right) \longrightarrow \Omega
_{i+j}\left( \mathcal{A},\mathcal{G}\right)
\end{equation}
compatible with $\delta ^{\mathcal{G}}$ with The multiplication is given by
the same table.
\end{remark}

We construct a canonical complex $\Omega \left( \mathcal{A},\mathcal{G},%
\mathcal{H}\right) $ over $\left( \mathcal{A},\mathcal{G},\mathcal{H}\right) 
$ as follows [14] 
\begin{equation}
\forall i=0,1,2,3:\Omega _{i}\left( \mathcal{A},\mathcal{G},\mathcal{H}%
\right) =\Omega _{i}\left( \mathcal{A},\mathcal{G}\right)
\end{equation}
and we define the exterior derivative $\delta $ on $\Omega \left( \mathcal{A}%
,\mathcal{G},\mathcal{H}\right) $ as follows 
\begin{eqnarray*}
\delta &=&\delta ^{\mathcal{G}}\text{ on }\Omega _{i}\left( \mathcal{A},%
\mathcal{G}\right) \text{ for }i=0,2,3 \\
\delta \left( g\otimes X+h\otimes Y\right) &=&\delta ^{\mathcal{G}}\left(
g\otimes X+h\otimes Y\right) -\delta ^{\mathcal{H}}\left( g\otimes
X+h\otimes Y\right) \text{ on }\Omega _{1}\left( \mathcal{A},\mathcal{G},%
\mathcal{H}\right)
\end{eqnarray*}
Before acting $\delta ^{\mathcal{H}}$ on \ $g\otimes X+h\otimes Y$, we do
the projection this element on $\mathcal{H\otimes A}$. The product $\ast $
in $\Omega \left( \mathcal{A},\mathcal{G},\mathcal{H}\right) $ is defined by 
\begin{eqnarray}
\ast &=&\ast _{\mathcal{G}}\text{ on }\Omega _{i}\left( \mathcal{A},\mathcal{%
G}\right) \text{ for }i=0,2,3, \\
\ast &=&\ast _{\mathcal{G}}-\ast _{\mathcal{H}}\text{ on }\Omega _{1}\left( 
\mathcal{A},\mathcal{G},\mathcal{H}\right)
\end{eqnarray}
The product $\ast $ and the coboundary $\delta $ verify the Leibniz rule.

\begin{proposition}
We can also construct a modified complex $\widetilde{\Omega }\left( \mathcal{%
A}^{\prime },\mathcal{G},\mathcal{H}\right) $ on $\mathcal{A}^{\prime }$ in
the following way 
\begin{eqnarray*}
\widetilde{\delta } &=&\widetilde{\delta }^{\mathcal{G}}\text{ on }%
\widetilde{\Omega }_{i}\left( \mathcal{A}^{\prime },\mathcal{G}\right) \text{
for }i=0,2,3 \\
\widetilde{\delta } &=&\widetilde{\delta }^{\mathcal{G}}-\widetilde{\delta }%
^{\mathcal{H}}\text{ on }\widetilde{\Omega }_{1}\left( \mathcal{A}^{\prime },%
\mathcal{G},\mathcal{H}\right)
\end{eqnarray*}
Before acting $\delta ^{\mathcal{H}}$ on an 1-form, we do the projection of
this element on $\mathcal{H\otimes A}^{\prime }$. The product $\ast $ in $%
\widetilde{\Omega }\left( \mathcal{A}^{\prime },\mathcal{G},\mathcal{H}%
\right) $ is defined by 
\begin{eqnarray}
\widetilde{\ast } &=&\widetilde{\ast }_{\mathcal{G}}\text{ on }\widetilde{%
\Omega }_{i}\left( \mathcal{A}^{\prime },\mathcal{G}\right) \text{ for }%
i=0,2,3, \\
\widetilde{\ast } &=&\widetilde{\ast }_{\mathcal{G}}-\widetilde{\ast }_{%
\mathcal{H}}\text{ on }\widetilde{\Omega }_{1}\left( \mathcal{A}^{\prime },%
\mathcal{G},\mathcal{H}\right) .  \notag
\end{eqnarray}
\end{proposition}

\begin{proof}
To prove the previous two propositions, it is sufficient to prove it in the
case $\mathcal{A}^{\prime }\mathcal{=A}$ with $\mathcal{A}$ is as in the
definitions (3)(4). To illustrate this idea, we compute $\widetilde{\delta }%
^{\mathcal{G}}\circ \widetilde{\delta }^{\mathcal{G}}\left( g\otimes
X\otimes m\right) $ with $g\otimes X\otimes m\in \widetilde{\Omega }%
_{1}\left( \mathcal{A}^{\prime },\mathcal{G}\right) $, 
\begin{eqnarray*}
\widetilde{\delta }^{\mathcal{G}}\circ \widetilde{\delta }^{\mathcal{G}%
}\left( g\otimes X\otimes m\right) &=&\widetilde{\delta }^{\mathcal{G}%
}\left( \left\{ C_{1}^{\mathcal{G}},g\right\} \otimes \left\{ C_{2}^{%
\mathcal{G}},X\right\} \otimes m+\frac{1}{2}d_{\mathcal{G}}\left( g\otimes
X\otimes m\right) \right) \\
&=&\left( -1\right) ^{\left\{ C_{1}^{\mathcal{G}},g\right\} }e\otimes
\left\{ \left\{ C_{1}^{\mathcal{G}},g\right\} ,\left\{ C_{2}^{\mathcal{G}%
},X\right\} \right\} \otimes m \\
&&+\frac{1}{2}\left( -1\right) ^{g}d_{\mathcal{G}}e\otimes \left\{
g,X\right\} \otimes m \\
&=&\left[ \left( -1\right) ^{\left\{ C_{1}^{\mathcal{G}},g\right\} }e\otimes
\left\{ C_{1}^{\mathcal{G}}g,\left\{ C_{2}^{\mathcal{G}},X\right\} \right\} +%
\frac{1}{2}\left( -1\right) ^{g}d_{\mathcal{G}}e\otimes \left\{ g,X\right\} %
\right] \otimes m
\end{eqnarray*}
It is clear that is equivalent to prove the nilpotency of $\widetilde{\delta 
}^{\mathcal{G}}$ in the abelian case. We use the same trick to prove the
other assertions.
\end{proof}

\begin{corollary}
There is a simple relation between the complex in the abelian case and the
non abelian case 
\begin{eqnarray*}
\widetilde{\delta }^{\mathcal{G}}\left( e\otimes X\otimes m\right) &=&\delta
^{\mathcal{G}}(e\otimes X)\otimes m,\qquad e\otimes X\otimes m\in \Omega
_{0}, \\
\widetilde{\delta }^{\mathcal{G}}\left( g\otimes X\otimes m\right) &=&\delta
^{\mathcal{G}}\left( g\otimes X\right) \otimes m,\qquad g\otimes X\otimes
m\in \Omega _{1}, \\
\widetilde{\delta }^{\mathcal{G}}\left( k\otimes Y\otimes n\right) &=&\delta
^{\mathcal{G}}\left( k\otimes Y\right) \otimes n,\qquad k\otimes Y\otimes
n\in \Omega _{2}.
\end{eqnarray*}
and the product on 1-forms can be written 
\begin{eqnarray}
\left( e\otimes X\otimes m\right) \widetilde{\ast }_{\mathcal{G}}\left(
e\otimes Y\otimes n\right) &=&e\otimes XY\otimes mn \\
&=&\left( e\otimes X\right) \ast _{\mathcal{G}}\left( e\otimes Y\right)
\otimes mn
\end{eqnarray}
in the same way we can formally we note $\widetilde{\ast }_{\mathcal{G}%
}\equiv \ast _{\mathcal{G}}\otimes \times $, where $\times $ is the matrix
product.
\end{corollary}

\subsection{Modified complex on supersphere and on fuzzy supersphere}

We consider the previous complex on the fuzzy supersphere $\mathcal{A}_{N}$
(for $N=\{1,2,...,\infty \}$ for a particular choice of $\mathcal{G}=sl(2,1)$%
, $\mathcal{H}=ops(2,1)$. Recall the $sl(2,1)$ is generated by $R_{\pm
},R_{3},\Gamma ,V_{\pm },D_{\pm }$ and $osp(2,1)$ by $R_{\pm },R_{3},V_{\pm
} $. Thus we have

1) Abelian case 
\begin{eqnarray}
\Omega _{0} &=&\Omega _{3}=\mathcal{A}_{N} \\
\Omega _{1} &=&\Omega _{2}=\bigoplus_{i=0}^{8}\left( \mathcal{A}_{N}\right)
_{i}
\end{eqnarray}

2) Non abelian

\begin{eqnarray}
\widetilde{\Omega }_{0} &=&\widetilde{\Omega }_{3}=\mathcal{A}_{N}\otimes
M_{n}\left( \mathbb{C}\right) \\
\widetilde{\Omega }_{1} &=&\widetilde{\Omega }_{2}=\bigoplus_{i=0}^{8}\left( 
\mathcal{A}_{N}\otimes M_{n}\left( \mathbb{C}\right) \right) _{i}
\end{eqnarray}

In details, all the forms are written as follows using a basis of the $%
\mathcal{G}$ : 
\begin{equation}
\text{0-form}\equiv \phi \text{ }
\end{equation}
\begin{eqnarray}
\text{1-form} &\equiv &r_{+}\otimes C_{+}+r_{-}\otimes C_{-}+r_{3}\otimes
C_{3}+\gamma \otimes W, \\
&&+v_{+}\otimes A_{+}+v_{-}\otimes A_{-}+d_{+}\otimes B_{+}+d_{-}\otimes
B_{-}  \notag \\
&\equiv &\left( A_{+},A_{-},B_{+},B_{-},C_{+},C_{-},C_{3},W\right)  \notag
\end{eqnarray}
in the same way 
\begin{equation}
\text{2-form}\equiv \left( a_{+},a_{-},b_{+},b_{-},c_{+},c_{-},c_{3},w\right)
\notag
\end{equation}

All these elements are in $\mathcal{A}_{N}$ or in $\mathcal{A}_{N}\otimes
M_{n}\left( \mathbb{C}\right) $. In the second case $A_{+}=A_{+}^{i}E_{i}$
with $E_{i}$ a basis of $M_{n}\left( \mathbb{C}\right) $. And the Casimir
element $C^{\mathcal{G}}-C^{\mathcal{H}}$ in this basis is 
\begin{equation}
C=2d_{-}\otimes d_{+}-2d_{+}\otimes d_{-}-\frac{1}{2}\gamma \otimes \gamma
\end{equation}
or 
\begin{equation}
C=2d_{-}\otimes d_{+}\otimes \mathbb{I}_{n}-2d_{+}\otimes d_{-}\otimes 
\mathbb{I}_{n}-\frac{1}{2}\gamma \otimes \gamma \otimes \mathbb{I}_{n}
\end{equation}
In first we have to explicit the product between forms 
\begin{eqnarray*}
\phi ^{1}\widetilde{\ast }\phi ^{2} &=&\phi ^{1}\phi ^{2} \\
\phi \widetilde{\ast }\left( A_{\pm },W,C_{i},B_{\pm }\right) &=&\left( \phi
A_{\pm },\phi W,\phi C_{i}\phi ,B_{\pm }\right) \\
\phi \widetilde{\ast }\left( a_{\pm },w,c_{i},b_{\pm }\right) &=&\left( \phi
a_{\pm },\phi w,\phi c_{i},\phi b_{\pm }\right) \\
\phi \widetilde{\ast }\Phi &=&\phi \Phi \\
\left( A_{\pm },W,C_{i},B_{\pm }\right) \widetilde{\ast }\psi &=&\left( \psi
A_{\pm },\psi W,\psi C_{i},\psi B_{\pm }\right) \\
\left( A_{\pm }^{1},W^{1},C_{i}^{1},B_{\pm }^{1}\right) \widetilde{\ast }%
\left( A_{\pm }^{2},W^{2},C_{i}^{2},B_{\pm }^{2}\right) &=&\left( a_{\pm
}^{\prime },w^{\prime },c_{i}^{\prime },b_{\pm }^{\prime }\right)
\end{eqnarray*}
where 
\begin{eqnarray*}
\left( a_{\pm }^{\prime },w^{\prime },c_{i}^{\prime },b_{\pm }^{\prime
}\right)
&=&(W^{1}B_{+}^{2}-B_{+}^{1}W^{2}-2C_{+}^{1}A_{-}^{2}+2A_{-}^{1}C_{+}^{2}-2C_{3}^{1}A_{+}^{2}+2A_{+}^{1}C_{3}^{2},
\\
&&W^{1}B_{-}^{2}-B_{-}^{1}W^{2}-2C_{-}^{1}A_{+}^{2}+2A_{+}^{1}C_{-}^{2}-2C_{3}^{1}A_{-}^{2}+2A_{-}^{1}C_{3}^{2},
\\
&&-4B_{+}^{1}A_{-}^{2}+4B_{-}^{1}A_{+}^{2}-4A_{-}^{1}B_{+}^{2}+4A_{+}^{1}B_{-}^{2}
\\
&&-4A_{+}^{1}A_{+}^{2}, \\
&&2A_{-}^{1}A_{+}^{2}+2A_{+}^{1}A_{-}^{2}, \\
&&4A_{-}^{1}A_{-}^{2} \\
&&W^{1}A_{+}^{2}-A_{+}^{1}W^{2}, \\
&&W^{1}A_{-}^{2}-A_{-}^{1}W^{2});
\end{eqnarray*}
The product between 1-forms and 2-forms is written 
\begin{eqnarray*}
\left( A_{\pm },W,C_{i},B_{\pm }\right) \widetilde{\ast }\left( a_{\pm
},w,c_{i},b_{\pm }\right) &=&A_{\pm }a_{\pm }-A_{\pm }a_{\pm }+\frac{1}{4}Ww-%
\frac{1}{2}C_{+}c_{-} \\
&&--\frac{1}{2}C_{-}c_{+}-C_{3}c_{3}-B_{+}b_{-}+B_{-}b_{+}
\end{eqnarray*}
All other products vanish. Now we are ready to explicit the coboundary 
\begin{eqnarray*}
\widetilde{\delta }\phi &=&\left( D_{\pm }\phi ,\Gamma \phi ,R_{+},R_{-}\phi
,R_{3}\phi ,V_{\pm }\phi \right) \\
\widetilde{\delta }\left( A_{\pm },W,C_{i},B_{\pm }\right) &=&(\Gamma B_{\pm
}-V_{\pm }W-2R_{\pm }A_{\mp }+2D_{\mp }C_{\pm }\mp 2R_{3}A_{\pm }\pm 2D_{\pm
}C_{3} \\
&&+2A_{\pm },4V_{+}A_{-}-4V_{-}A_{+}+4D_{-}B_{+}-4D_{+}B_{-}+2W, \\
&&-4D_{+}A_{+}-C_{+}, \\
&&-C_{3}+2D_{-}A_{+}+2D_{+}A_{-}, \\
&&4D_{-}A_{-}-C_{-}, \\
&&-B_{\pm }-D_{\pm }W+\Gamma A_{\pm }); \\
\widetilde{\delta }\left( a_{\pm },w,c_{i},b_{\pm }\right)
&=&D_{+}a_{-}-D_{-}a_{+}+\frac{1}{4}\Gamma w-\frac{1}{2}R_{+}c_{-} \\
&&-\frac{1}{2}R_{-}c_{+}-R_{3}c_{3}-V_{+}b_{-}+V_{-}b_{+}.
\end{eqnarray*}
For example $\Gamma B_{\pm }$ means $\left( \Gamma B_{\pm }^{i}\right) E_{i}$
The action of the operators $R_{i},V_{\pm },D_{\pm },\Gamma $ is given in
(2.7)(2.14) and $\delta $ is $osp(2,1)$ invariant [14].

We say that the 1-form $V=\left( A_{\pm },W,C_{i},B_{\pm }\right) $
satisfies the reality condition $V^{\ast }=V$\ when we have 
\begin{eqnarray}
A_{+}^{\ast } &=&A_{-},\qquad A_{-}^{\ast }=-A_{+},\qquad B_{+}^{\ast
}=-B_{-},\qquad B_{-}^{-}=B_{+} \\
C_{+}^{\ast } &=&C_{-},\qquad C_{-}^{+}=C_{+},\qquad C_{3}^{-}=C_{3},\qquad
W^{\ast }=W.  \notag
\end{eqnarray}
in the non abelian case, the reality condition means we consider only
1-forms $\widetilde{V}=V\otimes h$ with $V=V^{\ast }$ and $h$ a hermitian $%
n\times n$ matrix.

\section{Fields theories}

\subsection{The noncommutative pure gauge and supersymmetric fields over $%
\mathcal{A}_{N}$}

The noncommutative pure supersymmetric electrodynamics\footnote{%
This case is studied in [14].} (respectively Yang-Mills) over $\mathcal{A}%
_{N}$ $\left( \text{resp. }\mathcal{A}_{N}^{\prime }\right) $is a theory of
1-forms in the complex $\Omega \left( \mathcal{A}_{N},\mathcal{G},\mathcal{H}%
\right) $ $\left( \text{resp. }\widetilde{\Omega }\left( \mathcal{A}%
_{N}^{\prime },\mathcal{G},\mathcal{H}\right) \right) $ defined by an action 
\begin{equation}
S(V)=\frac{1}{g^{2}}Trace\left[ STrace\left[ \alpha \triangleleft F%
\widetilde{\ast }F+\beta \left( V\widetilde{\ast }\widetilde{\delta }V+\frac{%
2}{3}V\widetilde{\ast }V\widetilde{\ast }V\right) \right] \right] .
\end{equation}
where $Trace$ is the usual trace on the matrices which is used in the non
abelian case, $F=\widetilde{\delta }V+V\widetilde{\ast }V$ is the field
strength of $V$, $\alpha $, $\beta $ are parameters, $g$ a coupling constant
and the Hodge triangle $\triangleleft $ is the identification map between $%
\widetilde{\Omega }_{1}\left( \mathcal{A}_{N},\mathcal{G},\mathcal{H}\right) 
$ and $\widetilde{\Omega }_{2}\left( \mathcal{A}_{N},\mathcal{G},\mathcal{H}%
\right) $. The connection $V$ is real 1-form $V^{\ast }=V$, verifying 
\begin{equation}
\left( \widetilde{\delta }V+V\widetilde{\ast }V\right) _{\mathcal{H}}=0,
\end{equation}
for the field theoretical application we need moreover constraint 
\begin{equation}
C\widetilde{\ast }V_{\mathcal{H}^{\perp }}+V_{\mathcal{H}^{\perp }}%
\widetilde{\ast }C+\frac{1}{N}\left( \triangleleft V_{\mathcal{H}^{\perp }}%
\widetilde{\ast }V_{\mathcal{H}^{\perp }}\right) =0.
\end{equation}
Constraint (4.2) implies that the theory contains only 1-forms only $V_{%
\mathcal{H}^{\perp }}$ constrained moreover by (4.3). We can write the
interaction with matter as follows [11]. 
\begin{equation}
S_{matter}(V)=STrace\left[ \left( \widetilde{\delta }^{\mathcal{G}}-%
\widetilde{\delta }^{\mathcal{H}}+V_{\mathcal{H}^{\perp }}\right) \Phi
^{\ast }\widetilde{\ast }\triangleleft \left( \widetilde{\delta }^{\mathcal{G%
}}-\widetilde{\delta }^{\mathcal{H}}+V_{\mathcal{H}^{\perp }}\right) \Phi 
\right]
\end{equation}

$S_{matter}(V)+S(V)$ gives the $\mathcal{H}$-supersymmetric\footnote{%
invariant with respect to $\mathcal{H}$-action.} Schwinger model (resp.
Yang-Mills theory) over $\mathcal{A}_{N}$ $\left( \text{resp.}\mathcal{A}%
_{N}^{\prime }\text{ }\right) $. This action (4.1) and constraints
(4.2)(4.3) are invariant by gauge transformation 
\begin{equation}
V\longrightarrow UVU^{-1}-\widetilde{\delta }UU^{-1},\qquad U\in \mathcal{U}%
\left( \mathcal{A}_{N}\otimes M_{n}\left( \mathbb{C}\right) \right) .
\end{equation}
and by\ action of $\mathcal{H}$. For the details of the $\mathcal{H}$-action
see [14].

In the abelian case, by the non commutativity of the algebra $\mathcal{A}%
_{N} $ we have the term $V\ast V$ but in the commutative case this term
disappears. In the non commutative case the operator $\delta $ commutes only
with elements of the form $U=\exp (i\rho )e$ where $e$ is the unit of $%
\mathcal{A}_{N}$. Thus the action (4.1) is the noncommutative deformation of
an $U(1)$ gauge theory.

In the non abelian case, using corollary (1) it is easy to show that $\delta 
$ commute with elements of $\mathcal{U}\left( \mathcal{A}_{N}\otimes
M_{n}\left( \mathbb{C}\right) \right) $. Thus the action (4.1) is also the
noncommutative deformation of an $U(n)$ gauge theory. Now we are going to
study commutative limit $N\longrightarrow \infty $ of (4.1) in the two cases

\subsubsection{Commutative abelian case}

In the case, we have a pure gauge field action with $V=\left( A_{\pm
},W,C_{\pm },C_{i},B_{\pm }\right) $ satisfying (3.58). 
\begin{equation}
S_{\infty }\left( V\right) =I\left[ \alpha ^{\prime }\delta V\ast
\triangleleft \delta V+\beta ^{\prime }V\ast \delta V\right] ,
\end{equation}
the constraint (4.2) becomes 
\begin{equation}
V_{\mathcal{H}^{\perp }}=\left( A_{+},A_{-},W,0,0,0,0,0\right)
\end{equation}
and (4.3) becomes 
\begin{equation}
d_{+}A_{-}-d_{-}A_{+}+\frac{1}{4}\gamma W=0
\end{equation}
It follows 
\begin{equation}
F\equiv \delta V=\left( F_{+},F_{-},f,0,0,0,0,0\right) \in
\bigoplus_{i=0}^{8}\left( \mathcal{A}_{\infty }\right) _{i}
\end{equation}
where $\alpha ^{\prime },\beta ^{\prime }$ are real parameters. The
constraints (4.8) gives the following constraints on the ''additional''
superfields $C_{\pm },C_{i},B_{\pm }$%
\begin{equation}
C_{\pm }=\mp D_{\pm }A_{\pm },\qquad C_{3}=2D_{-}A_{+}+2D_{+}A_{-},\qquad
B_{\pm }=-D_{\pm }W+\Gamma A_{\pm }.
\end{equation}
with a new parametrization 
\begin{equation}
A_{+}=\frac{1}{2}\left( A-\overline{z}\overline{A}\right) ,\qquad A_{-}=-%
\frac{1}{2}\left( zA+\overline{A}\right) ,\qquad W=\overline{b}\overline{A}%
-bA
\end{equation}
A long calculation gives us the following result obtained in [14]

\begin{lemma}
We have 
\begin{eqnarray}
F_{+} &=&-\frac{3}{2}\left[ \overline{z}\overline{D}\left( n\omega \right)
+D\left( n\omega \right) \right] -4d_{+}n\omega \\
F_{-} &=&\frac{3}{2}\left[ zD\left( n\omega \right) -\overline{D}\left(
\omega \right) \right] -4d_{-}n\omega \\
f &=&3\left[ bD(n\omega )+bD(n\omega )\right] -4\gamma n\omega
\end{eqnarray}
with $n=\left( 1+\overline{z}z+\overline{b}b\right) $, $\omega =\overline{D}%
A+D\overline{A}$.

The action (4.5) becomes 
\begin{equation}
S_{\infty }\left( V\right) =\frac{1}{2\pi i}\int d\overline{z}dzd\overline{b}%
db\left\{ \alpha \overline{D}(n\omega )D(n\omega )+\beta n\omega
^{2}\right\} .
\end{equation}
the parameters $\alpha $, $\beta $ are linear combinations of $\alpha
^{\prime }$, $\beta ^{\prime }$. This action is $osp(2,1)$ supersymmetric.
\end{lemma}

The gauge symmetry $A\longrightarrow A+iD\Lambda $, $\overline{A}%
\longrightarrow \overline{A}+i\overline{D}\Lambda $ gives the following
expressions for $A,\overline{A}$ by gauge fixation which eliminates some
components 
\begin{eqnarray}
iA &=&bv+\frac{1}{2}\overline{b}\frac{iu}{1+\overline{z}z}+\overline{b}%
b\left( \frac{\eta }{1+\overline{z}z}\right) \\
i\overline{A} &=&\frac{1}{2}b\frac{iu}{1+\overline{z}z}+\overline{b}%
\overline{v}+\overline{b}b\left( \frac{\overline{\eta }}{1+\overline{z}z}%
\right)
\end{eqnarray}
with $u$ real, $v$, $\overline{v}$ mutually complex conjugate and $\eta
^{\ast }=\overline{\eta }$. We obtain 
\begin{equation}
n\omega =iu+b\eta -\overline{b}\overline{\eta }+\overline{b}b\left( \left( 1+%
\overline{z}z\right) \left( \partial _{\overline{z}}v-\partial _{z}\overline{%
v}\right) +\frac{iu}{1+\overline{z}z}\right)
\end{equation}
To finish we obtain by taking $\alpha =-\beta $%
\begin{eqnarray*}
S_{\infty }\left( V\right) &=&\frac{-\alpha }{2\pi i}\int d\overline{z}%
dz\{-\left( 1+\overline{z}z\right) ^{2}\left( \partial _{\overline{z}%
}v-\partial _{z}\overline{v}\right) ^{2}+\partial _{\overline{z}}u\partial
_{z}u+\frac{u^{2}}{\left( 1+\overline{z}z\right) ^{2}} \\
&&+\eta \partial _{\overline{z}}\eta +\overline{\eta }\partial _{z}\overline{%
\eta }+2\frac{\overline{\eta }\eta }{\left( 1+\overline{z}z\right) }\}.
\end{eqnarray*}
Hence we conclude that the commutative limit of the (4.1) is indeed standard
supersymmetric Schwinger model on the ordinary sphere.

\subsubsection{Commutative non abelian case}

We consider the commutative limit of the action (4.1) and we obtain the pure
\ non abelian gauge field with $V$ satisfies the reality condition.

\begin{equation}
S_{\infty }\left( V\right) =\frac{1}{g^{2}}\int Trace\left[ \alpha ^{\prime
}\triangleleft F\ast F+\beta ^{\prime }\left( V\ast \delta V+\frac{2}{3}%
V\ast V\ast V\right) \right]
\end{equation}
with 
\begin{equation}
F=\delta V+V^{2}=\left( F_{+},F_{-},f,0,0,0,0,0\right) .
\end{equation}
Using (4.26 ) We shall go to give explicitly all the components of $F$. Let
us set

\begin{equation*}
V=\left( A_{\pm },W,C_{i},B_{\pm }\right)
\end{equation*}
constrained by (4.7) and (4.8). We have 
\begin{eqnarray*}
\delta V &=&(\Gamma B_{\pm }-V_{\pm }W-2R_{\pm }A_{\mp }+2D_{\mp }C_{\pm
}\mp 2R_{3}A_{\pm }\pm 2D_{\pm }C_{3} \\
&&+2A_{\pm },4V_{+}A_{-}-4V_{-}A_{+}+4D_{-}B_{+}-4D_{+}B_{-}+2W, \\
&&-4D_{+}A_{+}-C_{+}, \\
&&-C_{3}+2D_{-}A_{+}+2D_{+}A_{-}, \\
&&4D_{-}A_{-}-C_{-}, \\
&&-B_{\pm }-D_{\pm }W+\Gamma A_{\pm });
\end{eqnarray*}
and 
\begin{eqnarray*}
V^{2} &=&(\left[ W,B_{+}\right] +2\left[ A_{-},C_{+}\right] +2\left[
A_{+},C_{3}\right] , \\
&&\left[ W,B_{-}\right] +2\left[ A_{+},C_{-}\right] +2\left[ A_{-},C_{3}%
\right] , \\
&&-4\left\{ B_{+},A_{-}\right\} +4\left\{ A_{+},B_{-}\right\} , \\
&&-4A_{+}A_{+},2A_{-}A_{+}+2A_{+}A_{-},4A_{-}A_{-}, \\
&&\left[ W,A_{+}\right] ,\left[ W,A_{-}\right] );
\end{eqnarray*}
In components 
\begin{eqnarray*}
F\pm &=&(\Gamma B_{\pm }-V_{\pm }W-2R_{\pm }A_{\mp }+2D_{\mp }C_{\pm }\mp
2R_{3}A_{\pm }\pm 2D_{\pm }C_{3}+2A_{\pm } \\
&&+\left[ W,B_{\pm }\right] +2\left[ A_{\mp },C_{\pm }\right] +2\left[
A_{\pm },C_{3}\right] , \\
&&4V_{+}A_{-}-4V_{-}A_{+}+4D_{-}B_{+}-4D_{+}B_{-}+2W+4\left[ B_{-},A_{+}%
\right] +4\left[ A_{+},B_{-}\right] , \\
&&-4D_{+}A_{+}-C_{+}-4A_{+}A_{+}, \\
&&-C_{3}+2D_{-}A_{+}+2D_{+}A_{-}+2A_{-}A_{+}+2A_{+}A_{-}, \\
&&4D_{-}A_{-}-C_{-}+4A_{-}A_{-}, \\
&&-B_{\pm }-D_{\pm }W+\Gamma A_{\pm }+\left[ W,A_{\pm }\right] ,);
\end{eqnarray*}
The constraint (4.9) implies the following constraints on the ''extra''
super fields $C_{i},B_{\pm }$%
\begin{eqnarray}
C_{\pm } &=&\mp 4\left( D_{\pm }A_{\pm }+A_{\pm }A_{\pm }\right) \\
C_{3} &=&+2D_{-}A_{+}+2D_{+}A_{-}+2A_{-}A_{+}+2A_{+}A_{-}  \notag \\
B_{\pm } &=&-D_{\pm }W+\Gamma A_{\pm }  \notag
\end{eqnarray}
using (4.10), $\delta V$ and $V^{2}$ become

\begin{eqnarray*}
\delta V &=&(\Gamma \left( -D_{\pm }W+\Gamma A_{\pm }.\right) -V_{\pm
}W-2R_{\pm }A_{\mp }\mp 8D_{\mp }\left( \left( D_{\pm }A_{\pm }+A_{\pm
}A_{\pm }\right) \right) \mp 2R_{3}A_{\pm } \\
&&\pm 2D_{\pm }\left( 2D_{-}A_{+}+2D_{+}A_{-}+2\left(
A_{-}A_{+}+A_{+}A_{-}\right) \right) +2A_{\pm }, \\
&&4V_{+}A_{-}-4V_{-}A_{+}+4D_{-}\left( -D_{+}W+\Gamma A_{+}\right)
-4D_{+}\left( -D_{-}W+\Gamma A_{-}\right) +2W, \\
&&+4A_{+}A_{+}, \\
&&-2\left( A_{-}A_{+}+A_{+}A_{-}\right) , \\
&&-4A_{-}A_{-},0,0)
\end{eqnarray*}
and 
\begin{eqnarray*}
V^{2} &=&(\left[ W,\left( -D_{\pm }W+\Gamma A_{\pm }\right) \right] +2\left[
A_{\mp },\left( -4\left( D_{\pm }A_{\pm }+A_{\pm }A_{\pm }\right) \right) %
\right] \\
&&+2\left[ A_{\pm },2D_{-}A_{+}+2D_{+}A_{-}+2\left(
A_{-}A_{+}+A_{+}A_{-}\right) \right] , \\
&&-4\left\{ \left( -D_{+}W+\Gamma A_{+}\right) ,A_{-}\right\} +4\left\{
A_{+},\left( -D_{-}W+\Gamma A_{-}\right) \right\} , \\
&&-4A_{+}A_{+},2A_{-}A_{+}+2A_{+}A_{-},4A_{-}A_{-}, \\
&&0,0);
\end{eqnarray*}
Finally, we have 
\begin{eqnarray*}
F &=&(\Gamma \left( -D_{\pm }W+\Gamma A_{\pm }.\right) -V_{\pm }W-2R_{\pm
}A_{\mp }\mp 8D_{\mp }\left( \left( D_{\pm }A_{\pm }+A_{\pm }A_{\pm }\right)
\right) \mp 2R_{3}A_{\pm } \\
&&\pm 2D_{\pm }\left( 2D_{-}A_{+}+2D_{+}A_{-}+2\left(
A_{-}A_{+}+A_{+}A_{-}\right) \right) +2A_{\pm } \\
&&+(\left[ W,\left( -D_{\pm }W+\Gamma A_{\pm }\right) \right] +2\left[
A_{\mp },\left( -4\left( D_{\pm }A_{\pm }+A_{\pm }A_{\pm }\right) \right) %
\right] \\
&&+2\left[ A_{\pm },2D_{-}A_{+}+2D_{+}A_{-}+2\left(
A_{-}A_{+}+A_{+}A_{-}\right) \right] , \\
&&4V_{+}A_{-}-4V_{-}A_{+}+4D_{-}\left( -D_{+}W+\Gamma A_{+}\right)
-4D_{+}\left( -D_{-}W+\Gamma A_{-}\right) +2W \\
&&-4\left[ \left( -D_{+}W+\Gamma A_{+}\right) ,A_{-}\right] _{+}+4\left[
A_{+},\left( -D_{-}W+\Gamma A_{-}\right) \right] _{+}, \\
&&0,0,0,0,0)
\end{eqnarray*}
with 
\begin{eqnarray*}
F_{\pm } &=&\left( \Gamma ^{2}+2)\right) A_{\pm }-\left( \Gamma D_{\pm
}+V_{\pm }\right) W\mp 12D_{\mp }D_{\pm }A_{-}\pm 12D_{\pm }D_{\pm }A_{+} \\
&&\pm 4D_{\pm }\left( A_{-}A_{+}+A_{+}A_{-}\right) \mp 8D_{\mp }\left(
A_{\pm }A_{\pm }\right) \\
&&\mp 8\left[ A_{\mp },\left( D_{\pm }A_{\pm }\right) \right] \pm 4\left[
A_{\pm },D_{-}A_{+}+D_{+}A_{-}\right]
\end{eqnarray*}
\begin{eqnarray*}
f &=&2W+4\left( D_{+}D_{-}-D_{-}D_{+}\right) W+4\left( V_{+}-D_{+}\Gamma
\right) A_{-}-4\left( V_{-}-D_{-}\Gamma \right) A_{+} \\
&&+4\left[ D_{+}W-\Gamma A_{+},A_{-}\right] _{+}-4\left[ A_{+},D_{-}W-\Gamma
A_{-}\right] _{+}
\end{eqnarray*}
using the parametrization (4.11), we obtain 
\begin{eqnarray*}
F_{+} &=&-\frac{3}{2}\left[ \overline{z}\overline{D}\left( n\omega \right)
+D\left( n\omega \right) \right] -4d_{+}n\omega \\
&&-\frac{1}{2}\left( 1+z\overline{z}\right) D\left( A\overline{A}+\overline{A%
}A\right) -\frac{1}{2}\overline{z}\left( 1+z\overline{z}\right) \overline{D}%
\left( A\overline{A}+\overline{A}A\right) \\
&&+\frac{3}{2}\left( 1+z\overline{z}\right) \overline{D}A^{2}+\frac{3}{2}%
\overline{z}\left( 1+z\overline{z}\right) D\overline{A}^{2} \\
&&\overline{z}\left( 1+z\overline{z}\right) \left[ A,\overline{D}\overline{A}%
\right] -\frac{1}{2}\left( 1+z\overline{z}\right) \left[ A,D\overline{A}%
\right] \\
&&+\left( 1+z\overline{z}\right) \left[ \overline{A},DA\right] -\frac{1}{2}%
\overline{z}\left( 1+z\overline{z}\right) \left[ \overline{A},\overline{D}A%
\right] ,
\end{eqnarray*}
and 
\begin{eqnarray*}
F_{-} &=&\frac{3}{2}\left[ zD\left( n\omega \right) -\overline{D}\left(
\omega \right) \right] -4d_{-}n\omega \\
&&-\frac{1}{2}\left( 1+z\overline{z}\right) \overline{D}\left( A\overline{A}+%
\overline{A}A\right) +\frac{1}{2}z\left( 1+z\overline{z}\right) D\left( A%
\overline{A}+\overline{A}A\right) \\
&&-\frac{3}{2}\left( 1+z\overline{z}\right) z\overline{D}A^{2}+\frac{3}{2}%
\left( 1+z\overline{z}\right) D\overline{A}^{2} \\
&&+\left( 1+z\overline{z}\right) \left[ A,\overline{D}\overline{A}\right] +%
\frac{1}{2}z\left( 1+z\overline{z}\right) \left[ A,D\overline{A}\right] \\
&&-\frac{1}{2}\left( 1+z\overline{z}\right) \left[ \overline{A},\overline{D}A%
\right] -\left( 1+z\overline{z}\right) z\left[ \overline{A},DA\right] ,
\end{eqnarray*}
and 
\begin{equation*}
f=-2\left( 1+z\overline{z}\right) \left[ \overline{A},A\right] _{+}.
\end{equation*}
Thus the action (4.19) is written as

\begin{proposition}
\begin{equation*}
S=\int d\overline{z}dzd\overline{b}dbTrace\left[ \{\alpha \left( \overline{D}%
\left( n\widetilde{\omega }\right) +\left[ \overline{A},n\widetilde{\omega }%
\right] \right) \left( D\left( n\widetilde{\omega }\right) +\left[ A,n%
\widetilde{\omega }\right] \right) +\beta n\widetilde{\omega }^{2}\}\right]
\end{equation*}
where $\widetilde{\omega }=D\overline{A}+\overline{D}A+\left[ \overline{A},A%
\right] _{+}$ and $n=1+\overline{z}z+\overline{b}b$. The complex arbitrary
parameters $\alpha $, $\beta $ are linear combinations of $\alpha ^{\prime }$%
, $\beta ^{\prime }$.
\end{proposition}

\begin{corollary}
In the abelian case, we find out 
\begin{equation}
S=\int d\overline{z}dzd\overline{b}db\left[ Trace\{\alpha \overline{D}\left(
n\omega \right) D\left( n\omega \right) +\beta n\omega ^{2}\right]
\end{equation}
where 
\begin{equation}
\omega =D\overline{A}+\overline{D}A\text{ and }n=1+\overline{z}z+\overline{b}%
b
\end{equation}
\end{corollary}

\subsubsection{Supersymmetric invariance}

Now I will show that this action is supersymmetric invariant using the
infinitesimal action of $osp(2,1)$ on 1-forms[14] 
\begin{equation}
\Delta A=\left( \epsilon _{+}V_{+}+\epsilon _{-}V_{-}\right) A+\frac{1}{2}%
\epsilon _{-}bA
\end{equation}
\begin{equation}
\Delta \overline{A}=\left( \epsilon _{+}V_{+}+\epsilon _{-}V_{-}\right) 
\overline{A}+\frac{1}{2}\epsilon _{+}\overline{b}\overline{A}
\end{equation}
We note $\epsilon V\equiv \epsilon _{+}V_{+}+\epsilon _{-}V_{-}$ where $%
\epsilon _{\pm }$ are grassmanian parameters. And we recall that the
definition of the action of $V_{\pm }$ is following 
\begin{equation}
V_{\pm }A=\left\{ v_{\pm },A\right\}
\end{equation}
The definition of charge $v_{\pm }$ is given in (2.7). Thus we obtain the
following lemmas

\begin{lemma}
We found the variation of the following terms under $osp(2,1)$ infinitesimal
action 
\begin{eqnarray}
\Delta (n\omega ) &=&\epsilon V\left( n\omega \right) \\
\Delta (\left\{ A,\overline{A}\right\} ) &=&\epsilon V\left\{ A,\overline{A}%
\right\} \\
\Delta (\left\{ A,\overline{A}\right\} ) &=&\epsilon V\left\{ A,\overline{A}%
\right\} \\
\Delta (\left[ A,n\widetilde{\omega }\right] ) &=&\epsilon V\left( \left[ A,n%
\widetilde{\omega }\right] \right) +\frac{1}{2}\epsilon _{-}b\left[ A,n%
\widetilde{\omega }\right] \\
\Delta (\left[ \overline{A},n\widetilde{\omega }\right] ) &=&\epsilon
V\left( \left[ \overline{A},n\widetilde{\omega }\right] \right) -\frac{1}{2}%
\epsilon _{+}\overline{b}\left[ \overline{A},n\widetilde{\omega }\right]
\end{eqnarray}
and we obtain 
\begin{equation}
\Delta S=Trace\int \frac{d\overline{z}dzd\overline{b}db}{n}\epsilon V\left(
\{\alpha \left( \overline{D}\left( n\widetilde{\omega }\right) +\left[ 
\overline{A},n\widetilde{\omega }\right] \right) \left( D\left( n\widetilde{%
\omega }\right) +\left[ A,n\widetilde{\omega }\right] \right) +\beta n%
\widetilde{\omega }^{2}\}\right)
\end{equation}
\end{lemma}

\begin{lemma}
The following property holds in this framework 
\begin{equation}
\int \frac{d\overline{z}dzd\overline{b}db}{n}\epsilon V(f)=0
\end{equation}
with $f\in \mathcal{A}_{\infty }.$
\end{lemma}

\begin{proof}
Easy computation
\end{proof}

The supergauge symmetry $A\longrightarrow A+iD\Lambda +i[A,\Lambda ]$, $%
\overline{A}\longrightarrow \overline{A}+iD\overline{\Lambda }+i[\overline{A}%
,\Lambda ]$ with $\Lambda \in \mathcal{A}_{\infty }$, is also evident. We
recall the fields $A,\overline{A}$ are odd elements of the algebras $%
\mathcal{A}_{\infty }$ with values in a arbitrary Lie algebra ($u(n)$ or $%
su(n)$). It means that $A=A_{i}T^{i}$ with $T^{i}$ generators of the Lie
algebras and $A_{i}$ elements of the functions algebra on the supersphere.
After the gauge fixation which allows us to eliminate same components, we
have 
\begin{eqnarray}
iA &=&bv+\frac{1}{2}\overline{b}\frac{iu}{1+\overline{z}z}+\overline{b}b%
\frac{\eta }{1+\overline{z}z} \\
i\overline{A} &=&\frac{1}{2}b\frac{iu}{1+\overline{z}z}+\overline{b}v+%
\overline{b}b\frac{\overline{\eta }}{1+\overline{z}z}
\end{eqnarray}
which allow us to compute explicitly the action 
\begin{eqnarray*}
S &=&\frac{1}{2\pi i}\int Trace[-\alpha \left( \overline{\eta }\partial _{z}%
\overline{\eta }+\partial _{\overline{z}}u\partial _{z}u+\eta \partial _{%
\overline{z}}\eta \right) +\alpha \left( \left( 1+\overline{z}z\right)
^{2}\left( i\left[ v,\overline{v}\right] +\left( \partial _{\overline{z}%
}v-\partial _{z}\overline{v}\right) \right) ^{2}\right) \\
&&\alpha \frac{u^{2}}{\left( 1+\overline{z}z\right) ^{2}}+i\alpha \left( %
\left[ \overline{\eta },\overline{\eta }\right] _{+}v+\frac{1}{2}\frac{i}{1+%
\overline{z}z}\left[ \overline{\eta },\eta \right] _{+}u\right) \\
&&+i\alpha \left( \overline{v}\left[ \eta ,\eta \right] _{+}+\frac{1}{2}%
\frac{i}{1+\overline{z}z}u\left[ \overline{\eta },\eta \right] _{+}\right)
+2\beta \frac{u^{2}}{\left( 1+\overline{z}z\right) ^{2}}+2\beta \frac{%
\overline{\eta }\eta }{1+\overline{z}z} \\
&&+2\alpha iu\left( i\left[ v,\overline{v}\right] +\left( \partial _{%
\overline{z}}v-\partial _{z}\overline{v}\right) \right) +2\beta iu\left( i%
\left[ v,\overline{v}\right] +\left( \partial _{\overline{z}}v-\partial _{z}%
\overline{v}\right) \right) ]
\end{eqnarray*}
Certain terms of this action merit some explanations, we note the component
of the fields in an explicit way 
\begin{eqnarray*}
v &=&v_{i}T^{i},\qquad u=u_{j}T^{j},\qquad \overline{v}=\overline{v}_{h}T^{h}
\\
\eta &=&\eta _{k}T^{k},\qquad \overline{\eta }=\overline{\eta }_{g}T^{g}
\end{eqnarray*}
and the product is defined as follows

\begin{equation*}
uv=u_{j}v_{i}T^{i}T^{j}
\end{equation*}
where $u_{j}v_{i}$ is product in the function algebra on the supersphere and 
$T^{i}T^{j}$ is product in the enveloping algebra of the Lie algebra. For
example 
\begin{eqnarray*}
\left\{ \overline{\eta },\eta \right\} &=&\left\{ \eta _{k}T^{k},\overline{%
\eta }_{g}T^{g}\right\} \\
&=&\eta _{k}T^{k}\overline{\eta }_{g}T^{g}+\overline{\eta }_{g}T^{g}\eta
_{k}T^{k} \\
&=&\eta _{k}\overline{\eta }_{g}\left[ T^{k},T^{g}\right] \\
&=&\eta _{k}\overline{\eta }_{g}C_{kg}^{i}T_{i}
\end{eqnarray*}
with $C_{kg}^{i}$ constants structure of the Lie algebra. Now we explicit
the product of the type $\left\{ \overline{\eta },\overline{\eta }\right\} v$%
\begin{equation}
v\left\{ \overline{\eta },\overline{\eta }\right\} =v_{j}\overline{\eta }_{k}%
\overline{\eta }_{g}C_{kg}^{i}T_{j}T^{i}
\end{equation}
It is an element on the enveloping algebra. The parameters $\alpha ,\beta $
are arbitrary and choosing $\alpha =-\beta $, we obtain the Yang-Mills
action on the ordinary sphere with some extras mass terms as in the abelian
case [14]. The action is very close to the standard Yang-Mills in the flat
euclidean space. 
\begin{eqnarray}
S &=&\frac{-\alpha }{2\pi i}\int d\overline{z}dzTrace\{-\left( 1+\overline{z}%
z\right) ^{2}\left( i\left[ v,\overline{v}\right] +\left( \partial _{%
\overline{z}}v-\partial _{z}\overline{v}\right) \right) ^{2}+\partial _{%
\overline{z}}u\partial _{z}u \\
&&+\frac{u^{2}}{\left( 1+\overline{z}z\right) ^{2}}+\overline{\eta }\partial
_{z}\overline{\eta }+\eta \partial _{\overline{z}}\eta +2\frac{\overline{%
\eta }\eta }{1+\overline{z}z}  \notag \\
&&-iv\left[ \overline{\eta },\overline{\eta }\right] _{+}-iv\left[ \eta
,\eta \right] _{+}+\frac{1}{1+\overline{z}z}u\left[ \overline{\eta },\eta %
\right] _{+}\}.  \notag
\end{eqnarray}
In component the action (4.46), with the choice $Trace\left(
T_{i}T^{j}\right) =\delta _{ij}$, becomes 
\begin{eqnarray*}
S &=&\frac{-\alpha }{2\pi i}\int d\overline{z}dz\{-\left( 1+\overline{z}%
z\right) ^{2}\left( iC_{ij}^{k}v_{i}\overline{v}_{j}+\left( \partial _{%
\overline{z}}v_{k}-\partial _{z}\overline{v}_{k}\right) \right) ^{2} \\
&&+\frac{u_{i}^{2}}{\left( 1+\overline{z}z\right) ^{2}}+\overline{\eta }%
_{i}\partial _{z}\overline{\eta }_{i}+\eta _{i}\partial _{\overline{z}}\eta
_{i}+2\frac{\overline{\eta }_{i}\eta _{i}}{1+\overline{z}z} \\
&&-iC_{kg}^{j}v_{j}\overline{\eta }_{k}\overline{\eta }_{g}-iC_{mn}^{l}v_{l}%
\eta _{m}\eta _{n}+\frac{1}{1+\overline{z}z}C_{tx}^{s}u_{s}\overline{\eta }%
_{t}\eta _{x}\}.
\end{eqnarray*}

\section{Conclusions}

We have constructed the supersymmetric electrodynamics and Yang-Mills
theories on the noncommutative sphere using a modified differential complex
: These theories possess only finite number of degrees of freedom. They are
respectively supersymmetric and supergauge invariant such that these
commutative limits become the supersymmetric Schwinger model and
supersymmetric Yang-Mills on the\ ordinary sphere.

This is a new step towards the understanding of the role of the
noncommutative geometry in the nonperturbative regularization of QFT. The
supersymmetry approach allows us to consider scalars fields, gauge fields
and spinors fields in a canonical set-up and the supersymmetric and
supergauge invariance single the good constraints which give us the correct
theory.

\begin{acknowledgement}
I would like to thank my advisor Ctirad Klimcik for suggesting this problem
to me and for many enlightening discussions and comments. I am also grateful
to ''Soci\'{e}t\'{e} de secours des amis des sciences''.
\end{acknowledgement}


\begin{thebibliography}{99}
\bibitem{1}  M. LeBellac, \textit{Des ph\'{e}nom\`{e}nes critiques aux
champs de jauge}, InterEditions/Editions du CNRS 1988

\bibitem{2}  F. Berezin, \textit{General concept of quantization}, Comm.
Math. Phys. 40, 153-174 (1975)

\bibitem{3}  U. Carow-Watamura, S. Watamura, \textit{Differential Calculus
on Fuzzy Sphere and Scalar Field}, q-alg/9710034

\bibitem{4}  U. Carow-Watamura, S. Watamura, \textit{Noncommutative geometry
and gauge theory on fuzzy sphere}, hep-th/9801195

\bibitem{5}  M. Chaichian, A. Demichev, P. Presnajder, \textit{Quantum Field
Theory on Noncommutative Space-Times and the persistence of Ultraviolet
Divergences, }hep-th/9812180

\bibitem{6}  A. Connes, \textit{Noncommutative Geometry}, Academie Press,
new York, 1994.

\bibitem{7}  H. Grosse, C. Klimcik, P. Presnajder\textit{, Field Theory on a
Supersymmetric Lattice}, Comm. Math. Phys. 185, 155-175 (1997).

\bibitem{8}  H. Grosse, P. Presnajder, \textit{A treatment of the Schwinger
Model within Noncommutative geometry}, hep-th/9805085.

\bibitem{9}  H. Grosse, G. Reiter, \textit{The fuzzy Supersphere},
math-ph/9804013.

\bibitem{10}  H. Grosse, C. Klimcik, P. Presnajder, \textit{Topologically
Nontrivial Field Configurations in Noncommutative geometry}, hep-th/9510083.

\bibitem{11}  H. Grosse, J. Madore, \textit{A noncommutative version of the
Schwinger model}, Phys. lett. B283 (1992) 218.

\bibitem{12}  H. Grosse, P. Presnajder, \textit{The Dirac Operator on the
Fuzzy Sphere}, Lett. Math. Phys. 33 (1995) 171.

\bibitem{13}  E. Hawkins, \textit{Quantization of equivariant vector
bundles, q-alg/9708030.}

\bibitem{14}  \textit{C. }Klimcik\textit{, A nonperturbative regularization
of the supersymmetric Schwinger model}, Comm. Math. Phys. 206, 567-586
(1999), hep-th/9903112.

\bibitem{15}  C. Klimcik, \textit{Gauge theories on the non commutative
sphere}, Comm. Math. Phys. 199 (1998).

\bibitem{16}  C. Klimcik,\textit{\ Gauge field on the fuzzy sphere : Bosonic
warm-up}, unpublished.notes.

\bibitem{17}  C. Klimcik, \textit{An extended fuzzy supersphere and twisted
chiral superfields,} Comm. Math. Phys., hep-th/9903202.

\bibitem{18}  J. Madore, \textit{Fuzzy Physics}, Annals of Physics, Vol 219,
N$%
%TCIMACRO{\UNICODE[m]{0xb0}}%
%BeginExpansion
{{}^\circ}%
%EndExpansion
$1, (1992).

\bibitem{19}  J. Madore, \textit{The commutative limit of a matrix geometry}%
, J. Math. Phys. 32 (1991) 332.

\bibitem{20}  J. Madore, \textit{Fuzzy Sphere}, Class. Quant. Grav. 9 (1992)
155.

\bibitem{21}  J. Madore, \textit{Gravity on Fuzzy Space-Time}, gr-qc/9709002.

\bibitem{22}  J. Madore, \textit{An introduction to Noncommutative
Differential geometry and its Physical Applications}, Cambridge University
press, 2nd Ed. 1999.

\bibitem{23}  H. S. Snyder, \textit{Quantized space-time}, Phys. Rev. 71
(1947) 38.
\end{thebibliography}
\end{document}